\documentclass[10pt,journal,oneside,twocolumn]{IEEEtran}
\usepackage{rule-D}
\usepackage{booktabs}
\usepackage{mathrsfs}
\usepackage{amsmath}
\usepackage{amssymb}
\usepackage{subfigure}
\usepackage{epsfig}
\usepackage{amsfonts}
\usepackage{graphicx}
\usepackage{algorithm}
\usepackage{algorithmic}
\usepackage{cite}
\usepackage{stfloats}
\usepackage{pifont}
\usepackage{booktabs}
\usepackage{psfrag}
\usepackage{times}
\usepackage{moreverb}
\usepackage{setspace}
\usepackage{txfonts}
\usepackage{array}
\usepackage[figuresright]{rotating}
\usepackage{hyperref}  % Add bookmarks to PDF file
\usepackage{url}
\usepackage{verbatim}
\usepackage{textcomp}
\usepackage{color}
\makeatletter
\renewcommand{\maketag@@@}[1]{\hbox{\m@th\normalsize\normalfont#1}}%
\makeatother

\newtheorem{theorem}{Theorem}

\newtheorem{proof}{Proof}

\begin{document}
%Wireless Energy Transfer with Practical Non-linear Power Amplifier: Reconfigurable Holographic Surface or Phased Array?
\title{ 6D Movable Holographic Surface Assisted Integrated Data and Energy Transfer: A Sensing Enhanced Approach }

\author{Zhonglun~Wang,~\IEEEmembership{Graduate Student Member,~IEEE,}
	Yizhe~Zhao,~\IEEEmembership{Member, ~IEEE,}
	Gangming~Hu,
	Yali~Zheng,~\IEEEmembership{Member, ~IEEE,}~and
	Kun~Yang,~\IEEEmembership{Fellow, ~IEEE}

\thanks{Zhonglun Wang, Yizhe Zhao and Gangming Hu are with the School of Information and Communication Engineering, University of Electronic Science and Technology of China, Chengdu 611731, China, e-mail: 202021010524@std.uestc.edu.cn;
	  yzzhao@uestc.edu.cn; 202521010916@std.uestc.edu.cn

Yali Zheng and Kun Yang are with the State Key Laboratory of Novel Software Technology, Nanjing University, Nanjing, 210008, China, and School of Intelligent Software and Engineering, Nanjing University (Suzhou Campus), Suzhou, 215163, China, email: yalizheng@nju.edu.cn, kunyang@nju.edu.cn.

} 
}
\maketitle

\begin{abstract}
Reconfigurable holographic surface (RHS) enables cost-effective large-scale arrays with high spatial gain. However, its amplitude-controlled holographic beamforming suffers from directional fluctuations, making it difficult to fully exploit the spatial gain of RHS. Fortunately, the promising 6D movable antenna (6DMA) provides a potential solution to this problem.
In this paper, we study a 6D movable holographic surface (6DMHS) integrated data and energy transfer (IDET) system, where a three-stage protocol is proposed, consisting of an uplink sensing stage, an orientation adjustment stage and a downlink transmission stage, to coordinate the 6DMHS and effectively serve the IDET receivers. 
Firstly, the holographic-based sensing technology is proposed and the sensing information of the IDET receivers is exploited.
Secondly, by fixing the rotations with the sensing information,  the orientation optimization problem is formulated for designing the holographic beamforming of the RHS and adjusting the translations of the 6DMHS. As a result, the directions with maximum beamforming gain are aligned with each IDET receiver. 
Thirdly, by fixing the orientation of the 6DMHS and the holographic beamforming, the equivalent wireless channel is obtained. The IDET performance optimization problem is formulated for obtaining the optimal digital beamforming, power splitting factor and energy harvesting (EH) power.
Simulation results demonstrate that the proposed scheme is capable of improving the IDET performance compared to the benchmarks.
\end{abstract}


\section{Introduction}
\subsection{Backgrounds}
In the 6G era, the number of the Internet of Things (IoT) devices is expected to increase exponentially to support the Internet of Everything \cite{10892042,10430083,10663809}. Traditional power supply methods, such as batteries and wired grids, inevitably incur significant maintenance costs. To overcome this problem, the integrated data and energy transfer (IDET) technology has been proposed, enabling simultaneous wireless power supply and data transmission to the IoT devices. Moreover, the IDET receivers is anticipated to be widely deployed in the large-scale IoT networks, thereby significantly enhancing the network efficiency while reducing the maintenance overhead \cite{10599126}.

One of the most effective ways to enhance IDET performance is to employ massive phased arrays and design beamforming schemes that improve wireless transmission efficiency \cite{9454368}. 
However, the phased arrays exhibits certain inherent limitations. Specifically, the phased arrays require a lot of phase shifters, RF-chains and power amplifier, which may lead to a higher cost. Additionally, the complicated wiring and large physical size make it difficult to employ the massive phased array. Consequently, the cost and size may restrict the development of the traditional phased array \cite{7350112}.

Fortunately, the reconfigurable holographic surface (RHS), composed of the printed circuit board (PCB), DC control circuits, and diodes, has emerged as a promising transmit antenna technology for realizing massive MIMO and highly accurate beamforming \cite{9696209}. In particular, the RHS employs an amplitude-controlled holographic beamformer, where the holographic principle is leveraged to generate directional beamforming for wireless transmission. However, due to the absence of the imaginary component in holographic beamforming, the beamforming gain of the RHS exhibits directional variations, i.e., the gains in different directions are not identical. This motivates us to rotate the RHS such that its maximum beamforming direction aligns with the IDET receiver, thereby fully exploiting its beamforming gain and improving IDET performance. Moreover, the emerging six-dimensional movable antenna (6DMA), containing of three rotation dimensions and three translation dimensions, offers a promising opportunity to realize this scheme \cite{Shao20246DMA}. Hence, by integrating the RHS with the 6DMA, we propose a novel six-dimensional movable holographic surface (6DMHS) to exhaustively exploit the beamforming gain of the RHS and enhance IDET performance.

\subsection{Related Works}
% Related works of RHS
Extensive research efforts have been devoted to enhancing IDET performance. The related studies on IDET and 6DMHS are summarized as follows.
%To overcome the aforementioned limitations, the reconfigurable holographic surface (RHS) is adopted as an alternative to the traditional phased arrays, aiming to reduce the transmitter costs and enhance the spatial gain. The RHS consists of metamaterial radiating elements arranged with subwavelength spacing. Fabricated compactly on a printed circuit board (PCB), the RHS employs varactor diodes, direct current (DC) control circuits and other low-cost components. The operational principle of the RHS is detailed as follows.

Initially, the concept of the IDET is proposed in \cite{4595260}. Then, many methods are adopts for improving the IDET performance. Bruno \textit{et al.} \cite{8115220} proposed the dedicated OFDM-based IDET waveform to compensate the nonlinear energy harvesting (EH) circuit, thereby improving the IDET performance. Moreover, Kwon \textit{et al.} \cite{9454368} investigated the massive MIMO-assisted IDET transmitter and designing the beamforming to improve the IDET performance. Furthermore, to increase the coverage performance, Na \textit{et al.} \cite{11030755} studied the reconfigurable intelligent surface (RIS)-assisted IDET system and proved the existence of the RIS can improve the IDET performance effectively. 

The aforementioned works mainly adopted the phased array-assisted transmitter. 
To overcome the costs and size drawbacks of the traditional phased array, the RHS-assisted IDET transmitter was proposed.
The RHS was initially adopted for wireless communication. Deng \textit{eq al.} \cite{9696209} proposed the dedicated holographic beamforming for wireless communication, proving that the RHS can effectively improve the throughput performance. Then, for simplifying the design of the holographic beamforming, they \cite{9681843} proposed the holographic-pattern division multiple access (HDMA) technology, which outperformed the traditional space-division multiple access (SDMA) in terms of throughput. 
Moreover, to further improve the throughput, Di \textit{et al.} \cite{9393594} adopted the RHS-assisted transmitter in the wideband OFDM system and designed the holographic to against the beamsquaint effect, proving that RHS outperformed the traditional phased array in the wideband system.
For further improving the IDET performance, the RHS-assisted transmitter was also adopted in the IDET systems.
Azarbahram 	\textit{et al.} \cite{azarbahram2024waveformoptimizationbeamfocusing} jointly optimized the waveform and beamforming design in the RHS assisted wireless energy transfer (WET) system, demonstrating superior WET performance compared to the conventional phased arrays. Additionally, Huang \textit{et al.} \cite{10721321} optimized the holographic beamforming by considering the continuous-aperture RHS and practical electromagnetic-based wireless channels, thus attaining the performance limits achievable in the IDET systems. 

Simultaneously, the aforementioned works adopted fixed-position antenna (FPA) to achieve IDET, which may not exploit the  DoFs exhaustively. The movable antenna was proposed to address this problem. Firstly, the fluid antenna was proposed to exploit the higher spatial DoFs. Wong \textit{et al.} \cite{9264694} firstly  proposed  the fluid antenna-assisted communication systems, proving the effectiveness of this antenna. Then, Lin \textit{et al.} \cite{10980171} proposed the fluid antenna-assisted IDET systems, proving the improvement of the multiplexing gain for the fluid antenna. Note that the fluid antenna can only move in 2D space and the utilization of the DoFs is limited. To further improve the utilization of the DoFs, the 6DMA was proposed \cite{10752873}. Shao \textit{et al.} proposed the 6DMA-assisted system and designed the orientation of the 6DMA carefully, proving that the 6DMA was capable of achieving high DoFs and improving the IDET performance significantly.
Moreover, by considering the finite discrete rotations and positions, they \cite{Shao20246DMA} designed the optimal orientation and power allocation scheme for improving the wireless communication performance.
Furthermore, Wang \textit{et al.} \cite{111111} adopted the 6DMA in the IDET system, demonstrating that the 6DMA was capable of improving the IDET performance effectively.

\subsection{Motivations and Contributions}
However, the related works still have several drawbacks, which are summarized as follows.
\begin{itemize}
	\item For RHS-assisted IDET systems, none of the existing studies considered the directional property of holographic beamforming caused by the absence of its imaginary component, nor did they investigate its impact on IDET performance.
	\item For 6DMA-assisted IDET systems, none of the existing works considered equipping the system with large-scale antennas or RHS, which may limit further performance improvements. 
	\item For the holographic beamforming and the orientation optimization problems, the algorithm complexity scales with the square of the number of antennas, the number of rotations and the number of translations. When the antenna array or the RHS is sufficiently large, this complexity becomes prohibitively high, making the deployment of 6DMA in practice very challenging.
\end{itemize}

Motivated by these limitations, this paper proposes the 6DMHS-assisted IDET system. By considering the directional property of holographic beamforming, the RHS is rotated such that its maximum beamforming direction aligns with the IDET receiver to pursue higher performance. To address the prohibitive complexity of joint orientation and holographic beamforming design, a sensing-assisted scheme is further developed, while the sensing information is adopted to rotate all RHSs and design the holographic beamforming directly. As a result, the complexity may reduce significantly. Finally, a three-stage protocol is proposed to fully exploit the potential of 6DMHS in IDET systems. Our main contributions are summarized as follows.
\begin{itemize}
\item A 6DMHS-assisted IDET system is developed, in which the RHS is equipped with holographic-based sensing modules capable of performing sensing operations. The extracted sensing information
is utilized to adjust the spatial orientation of the 6DMHS, thereby aligning the direction of the maximum beamforming gain of the RHS with the IDET receivers to improve the IDET performance.
\item To enable high-accuracy holographic sensing, an FFT-based detection method is adopted to extract the
angular information of the receivers from the holographic image. Based on the sensing information,
an alternating optimization algorithm is proposed to jointly optimize the spatial orientation of the
6DMHS and the holographic beamforming of the RHS. Subsequently, with the 6DMHS orientation fixed, the
digital beamforming and power splitting factor are jointly optimized using fractional programming
techniques.
\item Simulation results validate the superiority of the proposed holographic-based sensing design over conventional sensing method. The 6DMHS-assisted transmitter achieves significantly better IDET performance than systems relying on fixed-position RHS. Moreover, the proposed design outperforms the idealized perfect CSI-based scheme by reducing pilot overhead while delivering improved overall
performance.
\end{itemize}

\section{System Model}
In this section, we first introduce the model of the 6DMHS-assisted IDET system. Then, we introduce the wireless channel model of the system. Finally, we propose the WDT and WET model of the system.

%In this section, we first propose the 6DMA-assisted RHS-based IDET system. Then, the sensing-assisted 6DMA protocol is introduced. Next, the power-detection-based holography sensing method is adopted, whereas the FFT-based detection method is proposed to extracted the sensing information. Finally, the IDET signal model is established. 

\subsection{6DMHS-Transmitter Model}
\begin{figure}[!ht]
	\centering
	{\subfigure[]{\includegraphics[width = 1\linewidth]{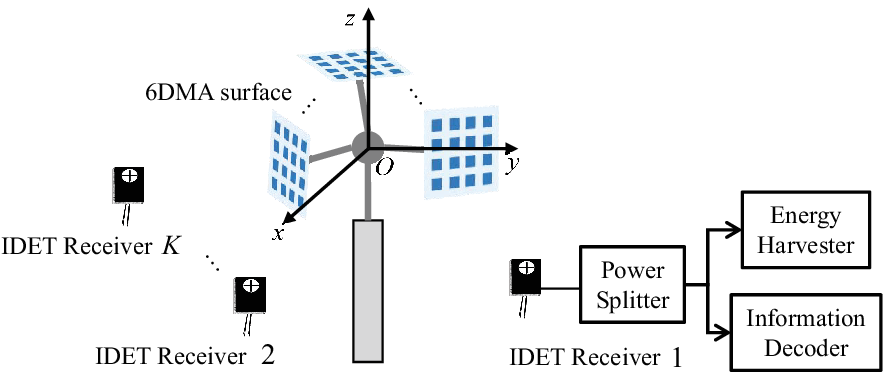}}
		\hfil
		\subfigure[]{\includegraphics[width = 1\linewidth]{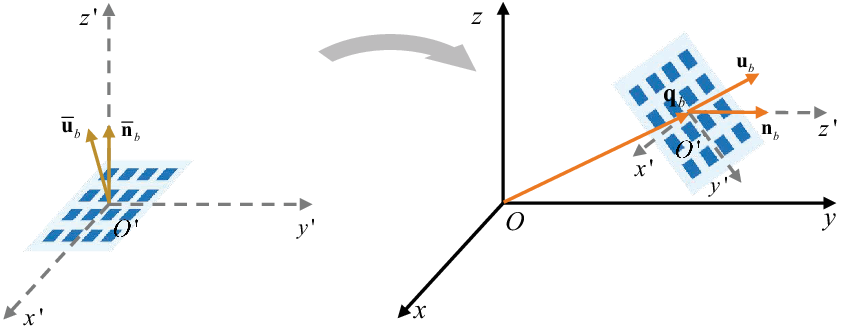}}
	}
	\setlength{\abovecaptionskip}{0pt}
	\setlength{\belowcaptionskip}{0pt}\caption{6DMHS model: (a) 6DMHS-based transmitter; (b) Illustration of the coordinate system of the 6DMHS.}
	\label{fig:1}
\end{figure}

As shown in Fig. \ref{fig:1}(a), the 6DMHS-transmitter consists of a base station and $B$ RHS. Each RHS is connected to the BS via a mechanical rod, along which a flexible cable is routed to transmit the IDET signals from the BS to the RHS. 
A translation motor and a rotary motor are installed at both ends of the mechanical rod to control its translation and rotation \cite{10752873}.
Moreover, each RHS is equipped with $Q$ feeds and $M = M_x \times M_y$ radiation elements, where $M_x$ and $M_y$ represent the number of the rows and columns of the RHS elements. The distance between two adjacent radiation elements is denoted as $d$. To fully utilize the beamforming gain of each RHS, we let each RHS serves one IDET receivers in a long frame. Therefore, there are $K=B$ single-antenna IDET receivers in the system. 

As shown in Fig. \ref{fig:1}(b), the RHS locted in the $x^\prime O^\prime y^\prime$ plane of the local coordinate system $O^\prime-x^\prime y^\prime z^\prime$. The coordinate of the $(m_x,m_y)$-th element in the local coordinate system is $\bar{\mathbf{r}}_{b,m_x,m_y} = [m_x d , m_y d , 0]^T \in \mathbb{R}^{3\times1}$ and the normal vector of the $b$-th RHS is $\bar{\mathbf{n}}_b \in \mathbb{R}^{3\times1}$. Moreover, $\bar{\mathbf{u}}_b \in \mathbb{R}^{3\times1}$ represents the maximal beamforming gain direction of the $b$-th RHS. 
At the global coordinate system $O-xyz$, the center point of the $b$-th RHS is $\mathbf{q}_b \in \mathbb{R}^{3\times1}$ and its normal vector is $\mathbf{n}_b \in \mathbb{R}^{3\times1}$. Moreover, the maximal beamforming gain direction of the $b$-th RHS at the global coordinate system is $\mathbf{u}_b \in \mathbb{R}^{3\times1}$.
The coordinate of the $(m_x,m_y)$-th element for the $b$-th RHS in the global coordinate system is given by
\begin{align}
	\mathbf{r}_{b,m_x,m_y} = \mathbf{q}_b + \mathbf{R}_b \overline{\mathbf{r}}_{b,m_x,m_y},
	\label{eq1}
\end{align}
where $\mathbf{R}_b \in \mathbb{R}^{3\times 3}$ is the unitary rotation matrix and is expressed as
\begin{align}
	\mathbf{R}_b = \cos \alpha_b \mathbf{I}_{3\times3} + \left( 1 - \cos\alpha_b  \right) \mathbf{v}_b \mathbf{v}_b^T + \sin \alpha_b \mathbf{V}_b. \label{eq2}
\end{align}
The matrix $\mathbf{V}_b \in \mathbb{C}^{3\times 3} $ is given as
\begin{align}
	\mathbf{V}_b = 
	\left[
	\begin{matrix}
		0 & -\mathbf{v}_b[3] & \mathbf{v}_b[2] \\
		\mathbf{v}_b[3] & 0 & -\mathbf{v}[1]   \\
		\mathbf{v}_b[2] & \mathbf{v}_b[1] & 0 
	\end{matrix}
	\right].\label{eq3}
\end{align}
Moreover, $\mathbf{v}_b$ and $\alpha_b$ represent the normalized rotation vector and the rotation angle of the $b$-th RHS, which can be expressed as 
\begin{align}
	&\mathbf{v}_b = \frac{ \bar{\mathbf{u}}_b \times  \mathbf{u}_b }{ \| \bar{\mathbf{u}}_b \times  \mathbf{u}_b \|_2 }, \label{eq4}\\
	&\alpha_b = \text{arccos} \frac{ (\bar{\mathbf{u}}_b)^T \mathbf{u}_b }{ \| \bar{\mathbf{u}}_b \|_2 \| \mathbf{u}_b \|_2 }, \label{eq5}
\end{align}
respectively.

According to Eq. \eqref{eq2}, we have
\begin{align}
	\mathbf{r}_{b,m_x,m_y} - \mathbf{q}_b = m_x d \mathbf{R}_b \mathbf{e}_1 +  m_y d \mathbf{R}_b \mathbf{e}_2,\label{eq6}
\end{align}	
where $\mathbf{e}_1 = [1,0,0]^T$ and $\mathbf{e}_2 = [0,1,0]^T$.

	Moreover, in order to ensure that the 6DMHS effectively radiates the IDET signal, three orientation constraints are introduced \cite{Shao2024DistributedCE}:
	\begin{itemize}
		\item \textbf{Rotation constraint for avoiding signal reflection:}
		\begin{align}
			\mathbf{n}_b^T ( q_{b_1} - q_{b_2} ) \leq 0,\ \forall b_1 \neq b_2. \label{eq7}
		\end{align}
		\item \textbf{Rotation constraint for avoiding signal blockage:}
		\begin{align}
			\mathbf{n}_b^T \mathbf{q}_b \geq 0, \forall b. \label{eq8}
		\end{align}
		\item  \textbf{Minimum-distance constraint for avoiding collision: }
		\begin{align}
			\| \mathbf{q}_{b_1} - \mathbf{q}_{b_2} \|_2 \geq d_\text{min}, \forall b_1 \neq b_2. \label{eq9}
		\end{align}
	\end{itemize}
The vector $\mathbf{n}_b$ denotes the outward normal of the $b$-th 6DMHS in the global coordinate system after translation and rotation operations.

%\begin{figure}
%	\centering
%	\includegraphics[width=1\linewidth]{figure/1-System_Model.eps}
%	\setlength{\abovecaptionskip}{0pt}
%	\setlength{\belowcaptionskip}{0pt} \caption{RHS-based 6DMHS transmitter.} \label{fig2}
%\end{figure}

%\begin{figure}
%	\centering
%	\includegraphics[width=0.75\linewidth]{figure/2-Coordinate.eps}
%	\setlength{\abovecaptionskip}{0pt}
%	\setlength{\belowcaptionskip}{0pt} \caption{Illustration of the coordinate system of the $b$-th RHS.} \label{fig30}
%\end{figure}

\subsection{Channel Model}

The wireless channel from the $k$-th IDET receiver to the $b$-th RHS is given as
\begin{align}
	\mathbf{h}_{k,b} =& \sqrt{M} \sum_\iota \Lambda_{k,\iota,b} \eta_{k,\iota,b} \mathbf{a}_{b}\left( \theta_{k,\iota},\phi_{k,\iota} \right), \label{eq10}
\end{align}
where $\theta_{k,\iota}$ and $\phi_{k,\iota}$ are the  azimuth and elevation angles of the $\iota$-the scatterer for the $k$-th receiver, respectively.
$\eta_{k,\iota,b}$ is the channel complex channel gain from the $k$-th IDET receiver to the $b$-th RHS through the $\iota$-th scatterer, while $\iota=0$ represents the LoS link and $\iota\geq 1$ represents the NLoS link. $\eta_{k,\iota,b}$ can be further expressed as
\begin{align}
	\eta_{k,\iota,b} = 
	\begin{cases}
		\sqrt{ \frac{K_\text{R}}{K_\text{R}+1} } \sqrt{\Omega_{k,b}}, \ \text{if}\ \iota = 0,\\
		\sqrt{ \frac{1}{K_\text{R}+1} } \sqrt{\Omega_{k,b}} \varkappa_{k,\iota,b}, \ \text{if}\ \iota \geq 1,
	\end{cases}
\end{align}
where $K_\text{R}$ is the Rician factor, $\Omega_{k,b}$ is the path-loss from the $k$-th IDET receiver to the $b$-th RHS and $\varkappa_{k,\iota,b} \sim \mathcal{CN}(0,1)$ is the complex channel gain from the $k$-th IDET receiver to the $b$-th RHS through the $\iota$-th scatterer.  $\Lambda_{k,\iota,b}$ is the antenna gain from the $k$-th IDET receiver to the $b$-th RHS through the $\iota$-th scatterer, which is given as \cite{10158988}
\begin{align}
	\Lambda_{k,\iota,b} = 
	\begin{cases}
		1,\ \text{if } \mathbf{q}_b^T \mathbf{f}(\theta_{k,\iota},\phi_{k,\iota}) > 0, \\
		0,\ \text{else}.
	\end{cases}
\end{align}
where the direction vector $\mathbf{f}(\theta,\phi)$ is defined as $\mathbf{f}(\theta,\phi) = \left[ f_1(\theta,\phi) , f_2(\theta,\phi) , f_3(\theta,\phi) \right]^T \in \mathbb{R}^{3\times1} $, where we have $f_1(\theta,\phi) = \cos\theta\cos\phi$, $f_2(\theta,\phi) = \sin\theta\cos\phi$ and $f_3(\theta,\phi) = \sin\phi$. The steering vector $\mathbf{a}_b\left( \theta_{k,\iota} , \phi_{k,\iota} \right) \in \mathbb{C}^{M\times 1}$ is defined as
\begin{align}
	\mathbf{a}_b\left( \theta,\phi \right) 
	=& \sqrt{\frac{1}{M}} \left[ e^{ j \frac{2\pi }{\lambda} \mathbf{f}^T(\theta,\phi) \mathbf{r}_{b,0,0} }, \cdots, e^{ j \frac{2\pi }{\lambda} \mathbf{f}^T(\theta,\phi) \mathbf{r}_{b,0,M_y-1} } , \cdots , \right. \nonumber\\
	&\ \ \ \ \left. e^{ j \frac{2\pi }{\lambda} \mathbf{f}^T(\theta,\phi) \mathbf{r}_{b,M_x-1,0} } , \cdots , e^{ j \frac{2\pi }{\lambda} \mathbf{f}^T(\theta,\phi) \mathbf{r}_{b,M_x-1,M_y-1} } \right]^T. \label{eq12}
\end{align}
Define the wireless channel from the 6DMHS-assisted transmitter to the $k$-th IDET receiver as $ \mathbf{h}_k = \left[ \mathbf{h}_{k,1}^T , \cdots , \mathbf{h}_{k,B}^T \right]^T \in \mathbb{C}^{BM\times1} $.

\subsection{Downlink Transmission Signal Model}
The downlink receive signal of the $k$-th IDET receiver is expressed as 
\begin{align}
	y_k &= \sum_{b} \sum_{k^\prime} \mathbf{h}_{k,b}^T \text{diag}(\mathbf{\Psi}_b) \boldsymbol{ \Theta }_b \mathbf{X}_{k^\prime,b} s_{k^\prime} + z_k,        
\end{align}
where $\mathbf{X}_{k,b}$ represents the beamforming vector for the $k$-th IDET receiver at the $b$-th RHS, $s_{k}$ is the IDET signal for the $k$-th receiver satisfying  $s_{k} \sim \mathcal{CN}(0,1)$ and $z_k$ is the antenna noise for the $k$-th receiver satisfying $z_k\sim\mathcal{CN}(0,\sigma_0^2)$.
$\boldsymbol{\Theta}_b$ represents the electromagnetic response of the $b$-th RHS and is denoted as $\boldsymbol{\Theta}_b = [ \boldsymbol{\Theta}_{b,0} , \cdots , \boldsymbol{\Theta}_{b,Q-1} ] \in \mathbb{C}^{M \times Q}$, where the $(m_xM_x+m_y)$-th entry of $\boldsymbol{\Theta}_{b,q} \in \mathbb{C}^{M\times1}$ is given as
\begin{align}
	{\Theta}_{b,q,m_xM_y+m_y} = \sqrt{\eta} \cdot e^{ -\alpha \| \mathbf{r}_{b,m_x,m_y} - \mathbf{r}_{b,q} \|_2 } \cdot e^{ -j \frac{ 2\pi f_c \varrho }{c} \| \mathbf{r}_{b,m_x,m_y} - \mathbf{r}_{b,q} \|_2 }, 
\end{align}
where $\eta$ is the efficiency factor, $\alpha$ is the real-valued propagation attenuation factor, $\varrho$ is the refractive factor of RHS and $\mathbf{r}_{b,q} \in \mathbb{R}^{3\times1}$ is the coordinate of the $q$-th feed at the $b$-th RHS, respectively.
Generally, the real attenuation term $e^{ -\alpha \| \mathbf{r}_{b,m_x,m_y} - \mathbf{r}_{b,q} \|_2 }$ is neglected, since $\alpha$ is a very small value \cite{9724245}.
$\boldsymbol{\Psi}_b \in \mathbb{C}^{M\times1}$ is the RHS beamforming vector for the $b$-th RHS.
%which is given as \cite{9848831}
%\begin{align}
%	\mathbf{\Psi}_b =  \sum_{q,k} \omega_{k,b,q} \frac{  \mathcal{R}\left\{ \sqrt{ \frac{M}{\eta} } \text{diag}(\boldsymbol{\Theta}_{b,q}) \mathbf{a}_b( \theta_{k}^\text{e} , \phi_k^\text{e} ) \right\} + 1  }{2}.
%\end{align}
%where $\omega_{k,b,q}$ is the weight factor. 

By employing the power splitter, the receive  signal at the $k$-th IDET receiver is divided into two parts: one for information decoding  and another for energy harvesting, which are expressed as
\begin{small}
	\begin{align}
		&y_{\text{ID},k} = \nonumber\\
		&\sqrt{1-\rho_k} \sum_{b} \sum_{k^\prime} \mathbf{h}_{k,b}^T \text{diag}(\mathbf{\Psi}_b) \boldsymbol{ \Theta }_b \mathbf{X}_{k^\prime,b}  s_{k^\prime} + \sqrt{1-\rho_k} z_k + z_{\text{cov},k},\\
		&y_{\text{EH},k} = \sqrt{\rho_k} \sum_{b} \sum_{k^\prime} \mathbf{h}_{k,b}^T \text{diag}(\mathbf{\Psi}_b) \boldsymbol{ \Theta }_b \mathbf{X}_{k^\prime,b} s_{k^\prime} + \sqrt{\rho_k} z_k,
	\end{align}
\end{small}where $\rho_k$ is the power splitting factor for the $k$-th receiver and $z_{\text{cov},k}$ is the passband-to-baseband conversion noise  satisfying $z_{\text{cov},k} \sim \mathcal{CN}(0,\sigma_\text{cov}^2) $ \cite{9795244}.

The SINR of the $k$-th IDET receiver is expressed as
\begin{small}
	\begin{align}
		&\gamma_k =\nonumber\\
		&\frac{ (1-\rho_k) \mathbb{E}\left[\left\vert \sum_{b} \mathbf{h}_{k,b}^T \text{diag}(\mathbf{\Psi}_b) \boldsymbol{\Theta}_b \mathbf{X}_{k,b} s_{k} \right\vert^2 \right] }{ (1-\rho_k) \mathbb{E} \left[ \left\vert \sum_b\sum_{k^\prime\neq k} \mathbf{h}_{k,b}^T \text{diag}(\mathbf{\Psi}_b) \boldsymbol{\Theta}_b \mathbf{X}_{k^\prime,b} s_{k^\prime} \right\vert^2 \right] + (1-\rho_k) \sigma_0^2 + \sigma_\text{cov}^2 }  \nonumber\\
		&= \frac{ (1-\rho_k) \left\vert \sum_{b} \mathbf{h}_{k,b}^T \text{diag}(\mathbf{\Psi}_b) \boldsymbol{\Theta}_b \mathbf{X}_{k,b}  \right\vert^2 }{ (1-\rho_k) \sum_{k^\prime\neq k} \left\vert \sum_b \mathbf{h}_{k,b}^T \text{diag}(\mathbf{\Psi}_b) \boldsymbol{\Theta}_b \mathbf{X}_{k^\prime,b} \right\vert^2  + (1-\rho_k) \sigma_0^2 + \sigma_\text{cov}^2 } 
	\end{align}
\end{small}Furthermore, the downlink throughput of the $k$-th IDET receiver is expressed as
\begin{align}
	R_k = \log_2\left( 1 + \gamma_k \right),\ [\text{bit/s/Hz}].
\end{align}

The EH power in terms of the RF signal for the $k$-th IDET receiver is expressed as
\begin{align}
	P_{\text{EH},k} &= \rho_k \mathbb{E}\left[ \left\vert \sum_{b} \sum_{k^\prime} \mathbf{h}_{k,b}^T \text{diag}(\mathbf{\Psi}_b) \boldsymbol{\Theta}_b \mathbf{X}_{k^\prime,b} s_{k^\prime} \right\vert^2 + \sigma_0^2 \right] \nonumber\\
	&= \rho_k  \sum_{k^\prime} \left\vert \sum_{b}  \mathbf{h}_{k,b}^T \text{diag}(\mathbf{\Psi}_b) \boldsymbol{\Theta}_b \mathbf{X}_{k^\prime,b}  \right\vert^2 + \rho_k \sigma_0^2.
\end{align}
After passing through the RF-to-DC conversion circuit, the harvested DC power for the $k$-th IDET receiver is expressed as
\begin{align}
	\Gamma( P_{\text{EH},k} ) = \max\left\{ \frac{E_\text{m}}{ e^{-\xi E_0 + \nu} } \left( \frac{ 1 + e^{ -\xi E_0 + \nu } }{ 1 + e^{-\xi P_{\text{EH},k}+\nu} } - 1 \right) , 0 \right\},
\end{align}
where $\xi$ and $\nu$ are the constant circuit parameters. $E_0$ and $E_\text{m}$ represent the activation and saturation power of the EH circuit, respectively.

\section{Directional Beamforming Gain of RHS and Protocol Design}

In this section, we first introduce the directional beamforming gain property of the RHS. Then, in view of this phenomenon, we propose the practical sensing-assisted protocol to adequately utilize the 6DMHS for IDET.

\subsection{Directional Beamforming Gain of RHS}

For the traditional phased array, the holographic beamformer is a complex vector and its phase varies over $[-\pi,\pi]$. However, for the RHS, the holographic beamformer $\boldsymbol{\Psi}_b$ is a real vector, meaning that the imaginary part of the beamformer vanishes. Hence, the RHS may not adjust the beamforming perfectly. 
We give an example as follows.

\begin{figure}
	%M = 128 * 32
	\centering
	\includegraphics[width=0.9\linewidth]{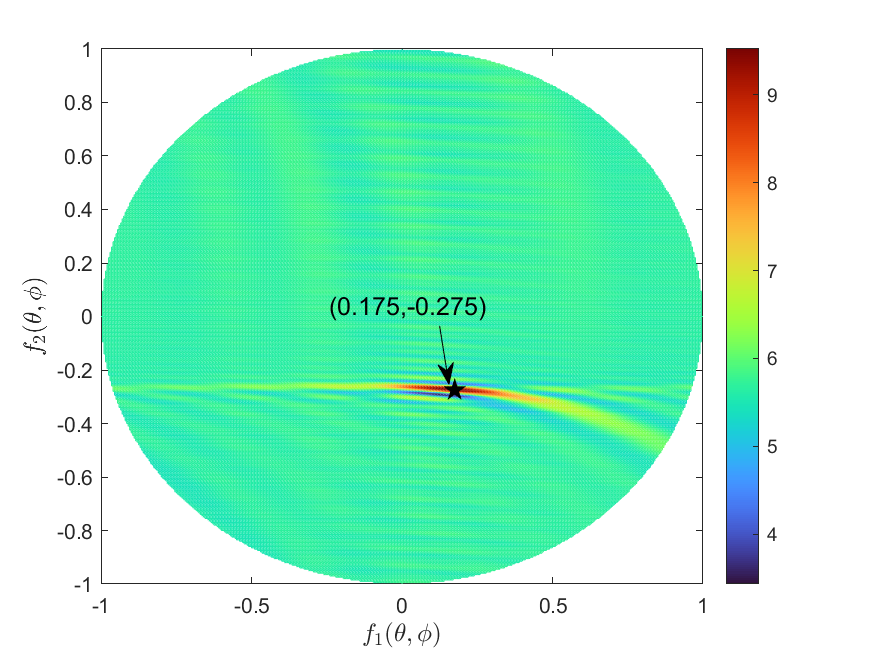}
	\setlength{\abovecaptionskip}{0pt}
	\setlength{\belowcaptionskip}{0pt} \caption{An illustration of the directional beamforming gain of the RHS.} \label{fig1}
\end{figure}

Denote the beamforming gain for the $q$-th feed of the $b$-th RHS towards the direction $\mathbf{f}(\theta,\phi)$ IDET receiver as $g_{b}(\theta,\phi)$, which is expressed as Eq. \eqref{BMGAIN} \cite{9393594,10163760,9848831}, where $ D_{b,q,m_x,m_y}$ and $\boldsymbol{\Psi}_{b}$ are given as 
\begin{align}
	&D_{b,q,m_x,m_y} = \left\| \mathbf{r}_{b,m_x,m_y} - \mathbf{r}_{b,q} \right\|_2. \nonumber\\
	&\mathbf{\Psi}_{b} =  \sum_{q} \omega_{b,q} \frac{  \mathcal{R}\left\{ \sqrt{ \frac{M}{\eta} } \text{diag}(\boldsymbol{\Theta}_{b,q}) \mathbf{a}_b( \theta , \phi ) \right\} + 1  }{2}.
\end{align}
Observe from Eq. \eqref{BMGAIN} that the beamforming gain consists of Parts I-III.
The Part I in Eq. \eqref{BMGAIN} contributes the mainly beamforming gain toward the direction $(\theta,\phi)$. The Part II in Eq. \eqref{BMGAIN} is the twin holographic beamforming gain towards the direction $(-\theta,-\phi)$ and it is generated due to the real-valued RHS beamforming. The Part III in Eq. \eqref{BMGAIN} is the interference  beamforming gain towards the other directions.
The Part II and Part III in Eq. \eqref{BMGAIN} may lead to a directional beamforming gain, \textit{i.e.}, the the beamforming gain is different at different directions. A demonstration of directional beamforming gain is shown in Fig. \ref{fig1}, where the maximum beamforming gain is achieved at the direction $\mathbf{f}(\theta,\phi) = [ 0.175 , - 0.275 , 0.945 ]^T$.
Consequently, if an IDET receiver is always at the maximum beamforming gain direction of the RHS, a higher IDET performance may be achieved.
In the next section, we adopt the 6DMA technique to adjust the orientation of the RHS for achieving the aforementioned method and enhancing the IDET performance. As a result, we give the definition of the directional beamforming gain of RHS.

\begin{figure*}[ht!]
	\begin{align}
		&g_{b}(\theta,\phi) = \left| \sum_q \mathbf{a}_b(\theta,\phi)^T \text{diag}(\boldsymbol{ \Psi }_b) \boldsymbol{\Theta}_{b,q} \right| \nonumber\\
		&= \frac{1}{\sqrt{M}} \left|    \sum_{q,q'} \omega_{b,q'} \sum_{m_x,m_y} e^{ -j\frac{2\pi}{\lambda}\left( \mathbf{f}^T(\theta,\phi) \mathbf{r}_{b,m_x,m_y} - \varrho D_{b,q,m_x,m_y} \right) }  \cdot \frac{ \mathcal{R}\left\{ e^{ j\frac{2\pi}{\lambda} \left( \mathbf{f}^T(\theta,\phi) \mathbf{r}_{b,m_x,m_y} - \varrho D_{b,q',m_x,m_y} \right) } + 1\right\}  }{2} \right| \nonumber\\
		&= \frac{1}{\sqrt{M}} \left| \underbrace{ \sum_{q} \frac{  \omega_{b,q^\prime } M }{4}  }_{ \text{Part I } }  + \underbrace{ \sum_{q}  \frac{\omega_{b,q}}{4} \sum_{m_x,m_y} e^{ -j\frac{2\pi}{\lambda}\left(  \mathbf{f}^T(\theta,\phi) \mathbf{r}_{b,m_x,m_y} - \varrho D_{b,q,m_x,m_y} \right) }  \cdot \frac{  e^{ -j\frac{2\pi}{\lambda} \left( \mathbf{f}^T(\theta,\phi) \mathbf{r}_{b,m_x,m_y} - \varrho D_{b,q,m_x,m_y} \right) } + 2  }{4} }_{ \text{ Part II } } \right. \nonumber\\
		&\ \ \ \  + \left. \underbrace{   \sum_{q\neq q^\prime} \omega_{b,q'} \sum_{m_x,m_y} e^{ -j\frac{2\pi}{\lambda}\left(  \mathbf{f}^T(\theta,\phi) \mathbf{r}_{b,m_x,m_y} - \varrho D_{b,q,m_x,m_y} \right) }  \cdot \frac{ \mathcal{R}\left\{ e^{ j\frac{2\pi}{\lambda} \left( \mathbf{f}^T(\theta,\phi) \mathbf{r}_{b,m_x,m_y} - \varrho D_{b,q',m_x,m_y} \right) } + 1\right\}  }{2} }_{ \text{ Part III } } \right| \label{BMGAIN}.
	\end{align}
	\rule{\linewidth}{1pt}
\end{figure*}

\textbf{ Definition 1 (Directional Beamforming Gain of RHS): } For an RHS with amplitude-controlled holographic beamforming, the beamforming gain is different for different direction $(\theta,\phi)$. Moreover, there exists the direction $(\theta^*,\phi^*)$,
which satisfies
\begin{align}
	 g_{b}(\theta^*,\phi^*) = \max_{ \substack{\theta,\phi,  \boldsymbol{\omega}_{b} \in \mathcal{S}_1,  \mathbf{r}_b^\text{feed} \in \mathcal{S}_2 }  }  g_{b}(\theta,\phi),\label{eq230}
\end{align}
where $\boldsymbol{\omega}_{b} = [ \omega_{b,0} , \cdots ,  \omega_{b,Q-1}  ]^T$ and $\mathbf{r}_b^\text{feed} = [ \mathbf{r}_{b,0} , \cdots , \mathbf{r}_{b,Q-1} ]$. Recall that the beamforming gain for the $q$-th feed of the $b$-th RHS towards the $k$-th IDET receiver $g_{b}(\theta^*,\phi^*)$ is defined as Eq. \eqref{BMGAIN}. The sets $\mathcal{S}_1$ and $\mathcal{S}_2$ are defined as 
\begin{align}
	\mathcal{S}_1 &= \left\{ (\boldsymbol{\omega}_b):  \boldsymbol{1}^T  \boldsymbol{\omega}_b = 1  \right\}, \\
	\mathcal{S}_2 &= \left\{ (\mathbf{r}_b^\text{feed}): \left\| \mathbf{r}_{b,i} - \mathbf{r}_{b,j}  \right\|_2^2 \geq \epsilon_r, \forall i\neq j \right\}, 
\end{align}
respectively. $\epsilon_r$ is the minimal distance of the adjacent feeds.

According to Eq. \eqref{eq230}, we can obtain the optimal feed positions and direction by adopting the Aquila Optimizer \cite{9950543}.

\textbf{Remark 1:} Once the RHS is fabricated, the feed positions are fixed. Therefore, in the subsequent sections, we adopt the optimal feed positions and keep them unchanged. Moreover, the direction  $(\theta,\phi)$ obtained from the Aquila Optimizer is defined as the maximum beamforming gain direction of the RHS, and is adopted when designing the orientation of the 6DMHS in the following part.

\subsection{Sensing-Enhanced 6DMHS Protocol}

\begin{figure}
	\centering
	\includegraphics[width=1\linewidth]{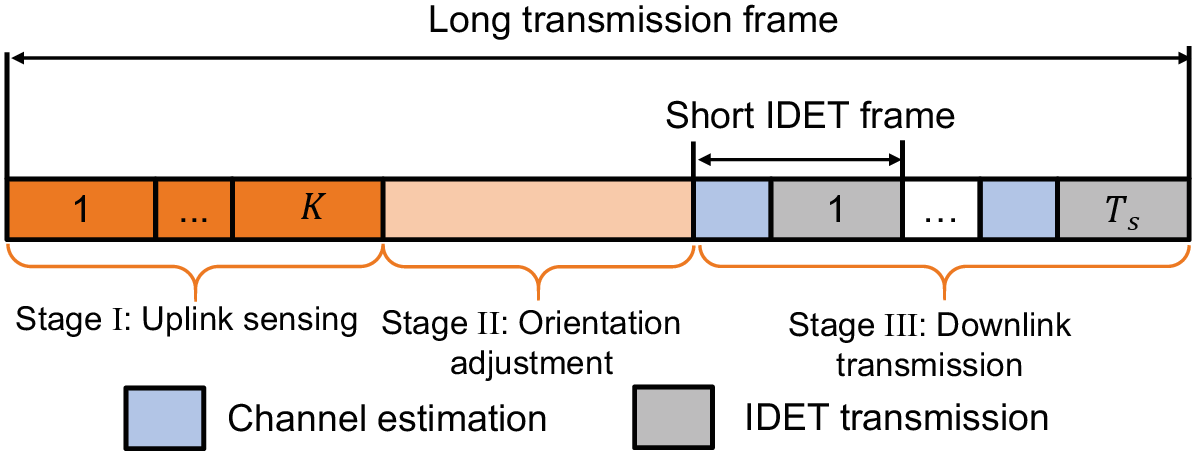}
	\setlength{\abovecaptionskip}{0pt}
	\setlength{\belowcaptionskip}{0pt} \caption{ Protocol of the 6DMHS-assisted IDET system.} \label{fig3}
\end{figure}

Denote the number of discrete positions of 6DMHS as $N$. Denote the number of discrete of rotations in each discrete position as $L$. To adjust the 6DMHS accurately, $N$ and $L$ are generally large enough. When adjusting the orientation of the 6DMHS, the algorithm complexity of the gradient-based algorithm for adopting the optimal scheme is $\mathcal{O}(M^2N^2L^2)$. For RHS, the number of elements is sufficiently large, which leading to a very high algorithm complexity. This indicates that it is difficult to employ the 6DMHS in practice. To address this problem, we first acquire the angle information by high accurate holographic-based sensing method, and then  align the RHS with the IDET receiver in the maximal beamforming gain direction for achieving higher WET performance by using the directional beamforming gain property of RHS.  The sensing-enhanced 6DMHS protocol is illustrated in Fig. \ref{fig3} and is summarized as follows.
\begin{itemize}
	\item Stage I (Uplink Sensing stage): All the RHS of 6DMHS are positioned at their initial locations. The IDET receivers sequentially transmit the sensing signals to the 6DMHS-assisted transmitter using the time division multiple access (TDMA) protocol. The BS then estimates the angles of the IDET receivers using the holographic-based sensing method.
	\item Stage II (Orientation adjustment): The 6DMHS-assisted transmitter jointly adjusts the orientation of the RHS and their holographic beamforming  based on the acquired sensing information, ensuring that the direction of the maximum beamforming gain of each RHS aligns with the corresponding sensing angle.
	\item  Stage III (Downlink transmission): At the beginning of each short IDET frame, the BS estimates the equivalent CSI of each receiver at each RF chain. This  equivalent CSI is defined as the cascade of the wireless channel, the holographic beamforming and the electromagnetic response of the RHS. Based on the estimated equivalent CSI, the digital beamforming and the power splitting strategy for each IDET receiver are then optimized. As a result, the IDET signals are transmitted to the IDET receivers successfully.
\end{itemize}

\section{Holographic-based Sensing Method and Orientation Adjustment for 6DMHS }

In this section, the holographic-based sensing method for RHS is first proposed. Then, by utilizing the sensing information, we propose the alternative optimization algorithm for adjusting the orientation of 6DMHS. 

\subsection{Holographic-based Sensing Method}

\begin{figure}
	\centering
	\includegraphics[width=1\linewidth]{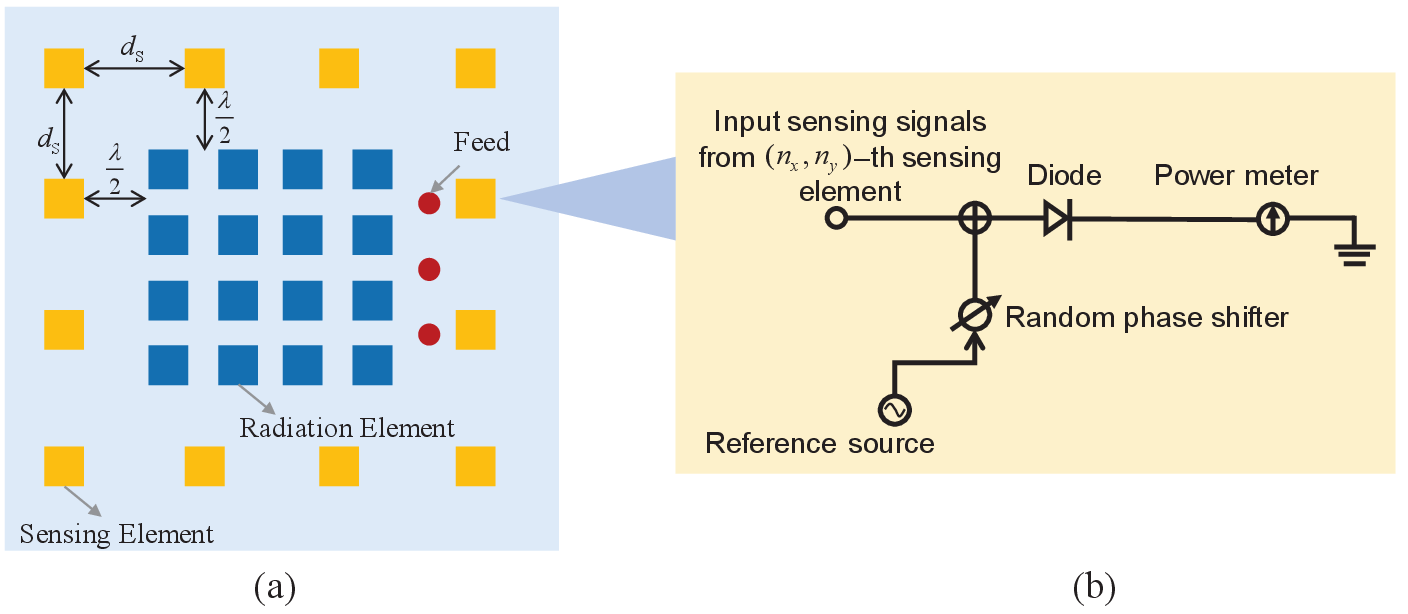}
	\setlength{\abovecaptionskip}{0pt}
	\setlength{\belowcaptionskip}{0pt} \caption{Illustration of the sensing RHS: (a) The architecture of the sensing RHS; (b) The circuit of the sensing element.} \label{fig40}
\end{figure}

For aligning the RHS with the IDET receiver accurately, a low-error angle information of the IDET receiver is required. However, the tradition sensing method may not acquire the sensing information accurately. The reasons are summarized as follows: 1) The loss of the imaginary part of the holographic beamformer; 2) In each feed, the receive sensing signal is the series superposition of the uplink sensing signal from each radiation element. Next, we propose the holographic-based sensing method for high accurate sensing.

The holography-based sensing model is illustrated in Fig. \ref{fig40}(a). The sensing elements are arranged around the RHS with the spacing of $d_\text{S}$. The holographic-based sensing circuit of each sensing element is illustrated as Fig. \ref{fig40}(b). When the uplink sensing signal is fed into the sensing circuit, it is superimposed with the reference signal. Then, this composite signal passes through the diode-based filter circuit, which retains only the direct current (DC) component. Next, the power meter records the DC power and uploads it to the BS. The collection of all recorded power values from all the power meters constitutes the holographic image, which encapsulates the CSI of all the IDET receivers.
The generation of the reference signals is summarized as follows: 1) Firstly, the reference source generates the signal with constant amplitude and phase; 2) Secondly, this signals pass through the random phase shifters, resulting in reference signals with randomized phases.
In the following, we describe how the holographic-based sensing is achieved by using the holographic image.

\textbf{Remark 2:} The sensing elements consist of the low-cost diode, reference source, low-precision random phase shifter and power meter, while they do not require the high-cost RF-chains and ADC modules \cite{10845870}. This indicates that the cost of the sensing elements is low and it is practical to equip them in the RHS for achieving higher sensing performance.

The wireless channel from the $k$-th IDET receiver to the sensing elements of the $b$-th RHS in the local coordinate system is given as 
\begin{align}
	&\mathbf{ H }_{k,b}^\text{S,L} = \nonumber\\
	&\left[ 
	\begin{array}{ccccc}
		{h}_{k,b,0,0}^\text{S,L} & {h}_{k,b,0,1}^\text{S,L} & \cdots & {h}^\text{S,L}_{k,b,0,N_y-2 } & {h}^\text{S,L}_{k,b,0,N_y-1 } \\
		{h}^\text{S,L}_{k,b,1,0} & 0 & \cdots & 0 & {h}^\text{S,L}_{k,b,1,N_y-1} \\
		\vdots & \vdots &  \ddots & \vdots & \vdots \\
		{h}^\text{S,L}_{k,b,N_x-2,0} & 0 & \cdots & 0 & {h}^{\text{S,L}}_{k,b,N_x-2,N_y-1}  \\
		{h}^\text{S,L}_{k,b,N_x-1,0} & {h}^\text{S,L}_{k,b,N_x-1,1} & \cdots & {h}^\text{S,L}_{k,b,N_x-1,N_y-2} & {h}^\text{S,L}_{k,b,N_x-1,N_y-1}
	\end{array}
	\right] ,
\end{align}
where $h_{k,b,n_x,n_y}^\text{S,L}$ is expressed as
\begin{align}
	h^\text{S,L}_{k,b,n_x,n_y} =
	\begin{cases}
	 \sum_\iota \Lambda_{k,\iota,b} \eta_{k,\iota,b} e^{ j \frac{2\pi}{\lambda} \mathbf{f}^T(\theta_{k,\iota,b}^\text{L},\phi_{k,\iota,b}^\text{L}) \mathbf{r}^\text{S,L}_{b,n_x,n_y} },\ \text{if}\ n_x \cdot n_y = 0\ \text{or}\ \\
	 \ \ \ \ \ \ \ \ \ \ \ \ \ \ \ \ \ \ \ \ \   n_x = N_x-1 \ \text{or}\  n_y = N_y-1, \\
	 0,\ \text{else},
	 \end{cases}
\end{align}
where $(\theta_{k,\iota,b}^{\text{L}}, \phi_{k,\iota,b}^{\text{L}})$ denote the angles of the $\iota$-th scatterer associated with the $k$-th IDET receiver in the local coordinate system of the $b$-th RHS.
The vector $\mathbf{r}^\text{S,L}_{b,n_x,n_y} \in \mathbb{R}^{3\times 1}$ represents the coordinates of the $(n_x,n_y)$-th sensing element at the $b$-th RHS in the local coordinate system.
The uplink receive signal from the $k$-th IDET receiver at the $(n_x,n_y)$-th sensing element of the $b$-th RHS is given as
\begin{align}
	y_{k,b,n_x,n_y}^\text{S} = \sqrt{P^\text{S}} h^\text{S,L}_{k,b,n_x,n_y}  s_k^\text{S} + z^\text{S}_{k,b,n_x,n_y},
\end{align}
where $P^\text{S}$ is the transmit power of the uplink sensing signal, $s_k^\text{S} = 1$ is the sensing signal and $z^\text{S}_{k,b,n_x,n_y} \sim \mathcal{CN}( 0 , \sigma_s^2 )$ is the antenna noise.

After being superimposed with the reference signal in the holographic-based sensing circuit, the output value of the $(n_x,n_y)$-th power meter at the $b$-th RHS is expressed as
\begin{align}
	&P^\text{S}_{k,b,n_x,n_y} \nonumber\\
	&= \left\vert y_{k,b,n_x,n_y}^{\text{S}} + s^{ \text{ref} }_{k,b,n_x,n_y} \right\vert^2 \nonumber \\
	&= \left\vert y_{k,b,n_x,n_y}^{\text{S}} \right\vert^2 + \left\vert s^{ \text{ref} }_{k,b,n_x,n_y} \right\vert^2 + y_{k,b,n_x,n_y}^{\text{S}*} s^{ \text{ref} }_{k,b,n_x,n_y} + y_{k,b,n_x,n_y}^{\text{S}} s^{ \text{ref}* }_{k,b,n_x,n_y} \nonumber \\
	&\overset{\text{(a)}}{\approx} \left\vert s^{ \text{ref} }_{k,b,n_x,n_y} \right\vert^2 + 2 
	\mathcal{R}\left\{ {y}_{k,b,n_x,n_y}^{\text{S}*} s^{ \text{ref} }_{k,b,n_x,n_y}\right\} \label{eq29}
\end{align}
where $s^{ \text{ref} }_{k,b,n_x,n_y} $  represents the reference signal of the $(n_x,n_y)$-th sensing element at the $b$-th RHS.
We have $s^{ \text{ref} }_{k,b,n_x,n_y} = \sqrt{A} e^{j2\pi \chi_{k,b,n_x,n_y}} $, where $A$ is a constant and   $\chi_{k,b,n_x,n_y} \sim \mathcal{U}( -\sigma_1 , \sigma_1 ) $.
In addition, (a) holds due to $\left\vert y_{k,b,n_x,n_y}^{\text{S}} \right\vert \ll \left\vert s^{ \text{ref} }_{k,b,n_x,n_y} \right\vert $ \footnote{ Note that the power of the reference wave can be artificially adjusted, whereas the receive signal power  is inherently low due to the limited uplink transmit power and the severe exponential path loss \cite{10845870}. As a result, the  reference wave power is significantly higher than that of the uplink signal. }.
Note that the reference signal is known at the BS. By subtracting the reference signal from Eq. \eqref{eq29}, the holographic image is given as
\begin{align}
	\tilde{P}^\text{S}_{k,b,n_x,n_y} = 2  \mathcal{R}\left\{ y_{k,b,n_x,n_y}^{\text{S}*} s^{ \text{ref} }_{k,b,n_x,n_y} \right\}.
\end{align}
Then, by exciting the holographic image with the reference signal, the angle information is extracted. The excited holographic image is expressed as
\begin{align}
	\mathcal{H}_{k,b,n_x,n_y} 
	&= \tilde{P}^\text{S}_{k,b,n_x,n_y} s^{ \text{ref}  }_{k,b,n_x,n_y} \nonumber\\
	&=A \sqrt{P^\text{S}} h^\text{S,L}_{k,b,n_x,n_y} + \sqrt{P^\text{S}} h_{k,b,n_x,n_y}^{\text{S,L}*} \left( s^{ \text{ref}  }_{k,b,n_x,n_y}\right)^2 \nonumber\\
	&\ \ \  + A z_{k,b,n_x,n_y}^\text{S} + z_{k,b,n_x,n_y}^\text{S*} \left( s^{ \text{ref}  }_{k,b,n_x,n_y}\right)^2. \label{eq20}
\end{align}

\begin{theorem}
	When the number of the sensing elements is sufficient large, \textit{i.e.}, $N_x,N_y\rightarrow \infty$, we have
		\begin{align}
			\tilde{\mathcal{H}}_{k,b}  
			&= \frac{1}{N} \sum_{n_x,n_y} e^{ -j \frac{2\pi}{\lambda} \mathbf{f}^T(\theta^\text{L,e}_{k,b},\phi^\text{L,e}_{k,b}) \mathbf{r}^\text{S,L}_{b,n_x,n_y} } \mathcal{H}_{k,b,n_x,n_y} \nonumber\\
			&= \frac{1}{N} \sum_{n_x,n_y} e^{ -j \frac{2\pi}{\lambda} \mathbf{f}^T(\theta^\text{L,e}_{k,b},\phi^\text{L,e}_{k,b}) \mathbf{r}^\text{S,L}_{b,n_x,n_y} } \left[ A\sqrt{P^\text{S}}  h^\text{S,L}_{k,b,n_x,n_y}   + \sqrt{P^\text{S}} 
			\right. \nonumber\\
			&\ \ \ \ \left.  \cdot h_{k,b,n_x,n_y}^{\text{S,L}*} \left( s^{ \text{REF}  }_{k,b,n_x,n_y}\right)^2 + A z_{k,b,n_x,n_y}^\text{S} + z_{k,b,n_x,n_y}^\text{S*} \left( s^{ \text{REF}  }_{k,b,n_x,n_y}\right)^2 \right] \nonumber\\
			&\overset{N\rightarrow\infty}{=} 
			\begin{cases}
				\sqrt{P^\text{S}}A\Lambda_{k,\iota,b}\eta_{k,\iota,b},\text{ if }f_1(\theta_{k,b}^\text{L,e},\phi_{k,b}^\text{L,e}) = f_1(\theta_{k,\iota,b}^\text{L},\phi_{k,\iota,b}^\text{L})\  \\ \ \ \ \ \ \ \ \ \ \ \ \ \ \ \ \ \text{and} \  f_2(\theta_{k,b}^\text{L,e},\phi_{k,b}^\text{L,e}) = f_2(\theta_{k,\iota,b}^\text{L},\phi_{k,\iota,b}^\text{L}),\\
				0,\text{ else},
			\end{cases}  \label{eq32}
		\end{align}
		where $N = 2N_x + 2 N_y - 4$. $(\theta_{k,b}^{\text{L,e}}, \phi_{k,b}^{\text{L,e}})$ denote the estimated angles of the $k$-th IDET receiver with respect to the $b$-th RHS.
\end{theorem}
\begin{proof}
	Please see Appendix A.
\end{proof}

According to \textit{Theorem 1}, the direction corresponding to the maximum beamforming gain for the $k$-th IDET receiver is given as 
\begin{align}
	( \theta_{k,b^*}^\text{L,e*} , \phi_{k,b^*}^\text{L,e*} , b^* ) = \arg\max_{b,\theta_{k,b}^\text{L,e},\phi_{k,b}^\text{L,e}} \tilde{\mathcal{H}}_{k,b}. \label{eq33}
\end{align}
By adopting the FFT-based detection method, Eq. \eqref{eq33} can be efficiently solved with  low computational complexity.

\begin{theorem}
	Densely arranged antennas do not improve the angular resolution of the RHS. Therefore, a half-wavelength spacing represents the optimal element configuration when using the FFT-based detection method.
\end{theorem}
\begin{proof}
	Please see Appendix B.
\end{proof}

According to \textit{Theorem 2}, it is not necessary to install the sensing module densely. Instead, we can place the sensing elements with a half-wavelength spacing, which helps reduce costs and simplifies the wiring complexity. 

Denote the matrix of the holographic image as 
\begin{small}
\begin{align}
	&\boldsymbol{ \mathcal{H} }_{k,b} = \nonumber\\
	&\left[ 
	\begin{array}{ccccc}
		\mathcal{H}_{k,b,0,0} & \mathcal{H}_{k,b,0,1} & \cdots & \mathcal{H}_{k,b,0,N_y-2 } & \mathcal{H}_{k,b,0,N_y-1 } \\
		\mathcal{H}_{k,b,1,0} & 0 & \cdots & 0 & \mathcal{H}_{k,b,1,N_y-1} \\
		\vdots & \vdots &  \ddots & \vdots & \vdots \\
		\mathcal{H}_{k,b,N_x-2,0} & 0 & \cdots & 0 & \mathcal{H}_{k,b,N_x-2,N_y-1}  \\
		\mathcal{H}_{k,b,N_x-1,0} & \mathcal{H}_{k,b,N_x-1,1} & \cdots & \mathcal{H}_{k,b,N_x-1,N_y-2} & \mathcal{H}_{k,b,N_x-1,N_y-1}
	\end{array}
	\right] .
\end{align}
\end{small}Then, by adopting the 2D-FFT, we have
\begin{align}
	(n_{x,b^*}^*,n_{y,b^*}^*,b^*) = \arg \max_{n_x,n_y,b} \left\vert \mathbf{F}_{N_x \times N_x}^T \boldsymbol{ \mathcal{H} } _{k,b} \mathbf{F}_{N_y \times N_y} \right\vert.\label{eq35}
\end{align}
According to \eqref{eq35}, the estimated direction vector in the local coordinate system is given as
\begin{align}
	&\mathbf{f}(\theta_{k,b^*}^\text{L,e},\phi_{k,b^*}^\text{L,e}) = \left[ \frac{(2n_{y,b^*}^*-N_y+1)d_\text{S}}{2N\lambda} , \frac{(2n_{x,b^*}^*-N_x+1)d_\text{S}}{2N\lambda} , \right. \nonumber\\
	&\left. \sqrt{ 1 - \left( \frac{(2n_{y,b^*}^*-N_y+1)d_\text{S}}{2N\lambda})^2 - (\frac{(2n_{x,b^*}^*-N_x+1)d_\text{S}}{2N\lambda} \right)^2 } \right]^T.
\end{align} 
Moreover, the estimated direction vector in the global coordinate system is given as
\begin{align}
	\mathbf{f}(\theta_{k}^\text{e},\phi_{k}^\text{e}) = \bar{\mathbf{R}}_{b^*} \mathbf{f}(\theta_{k,b^*}^\text{L,e},\phi_{k,b^*}^\text{L,e}),
\end{align}
where $\bar{\mathbf{R}}_{b^*}$ is the initial rotation matrix for the $b^*$-th RHS. Finally, we utilize the sensing information to adjust the orientation of the 6DMHS for improving the IDET performance.

\subsection{Orientation Adjustment for 6DMHS}

In our proposed protocol, the $b$-th RHS is adopted to serve the $b$-th IDET receiver.
The parameters $\bar{\mathbf{u}}_b$ and $\mathbf{u}_b$ are given as
\begin{align}
	\bar{\mathbf{u}}_b &= \left[ \cos \theta^* \cos \phi^* , \sin \theta^* \cos \phi^*, \sin \phi^* \right]^T, \label{eq38}\\
	\mathbf{u}_b & = \left[ \cos \theta_{k=b}^\text{e} \cos \phi_{k=b}^\text{e} , \sin \theta_{k=b}^\text{e} \cos \phi_{k=b}^\text{e} , \sin \phi_{k=b}^\text{e} \right]^T . \label{eq39}
\end{align}
According to Eqs.~\eqref{eq2}–\eqref{eq5}, \eqref{eq38}, and \eqref{eq39}, the unitary rotation matrix of the $b$-th RHS, denoted by $\tilde{\mathbf{R}}_b$, is obtained. In the following, $\tilde{\mathbf{R}}_b$ is employed to adjust the rotation of the $b$-th RHS.
The translation position of the $b$-th RHS $\mathbf{q}_b$ is given as
\begin{align}
	\mathbf{q}_b = \mathbf{s}_b^T \mathbf{q},
\end{align}
where $\mathbf{q}$ is the translation positions and is denoted as $\mathbf{q} = \left[ \mathbf{q}^{(0)} , \cdots , \mathbf{q}^{(M_1-1)} \right] \in \mathbb{R}^{M_1 \times 3} $. $\mathbf{q}^{(m_1)} \in \mathbb{R}^{3\times1}$ is the coordinate of the $m_1$-th translation position.
the selection vector $\mathbf{s}_b \in \mathbb{R}^{M_1 \times 1}$ is defined as
\begin{align}
	s_b[m_1] = 
	\begin{cases}
		1,\ \text{if}\ m_1=m_{1,b},\\
		0,\ \text{else}.
	\end{cases}
\end{align}
Here, the $m_{1,b}$-th position is selected for the $b$-th RHS. By substituting the obtained $\tilde{\mathbf{R}}_b$ and $\mathbf{q}_b$ into Eq. \eqref{eq12}, the steering vector $\mathbf{a}_b(\theta_{k=b}^\text{e},\phi_{k=b}^\text{e})$ is obtained.
Then, the holographic beamformer of the $b$-th RHS is expressed as
\begin{align}
	\bar{\boldsymbol{\Psi}}_b = \sum_{q} \bar{\omega}_{b,q}  \frac{  \mathcal{R} \left\{ \sqrt{\frac{M}{\eta}} \text{diag}\left( \boldsymbol{ \Theta }_{b,q} \right) \mathbf{a}_b(\theta_{k=b}^\text{e},\phi_{k=b}^\text{e}) \right\} + 1 }{ 2 }   .
\end{align}
The power gain of the $k$-th RHS at the estimated direction $(\theta_{k}^\text{e},\phi_{k}^\text{e})$ is given as
\begin{align}
	\dot{g}_k(\theta_{k}^\text{e},\phi_{k}^\text{e}) = \sum_{k^\prime} \left| \sum_b \mathbf{a}_b( \theta_{k}^\text{e} , \phi_{k}^\text{e}  ) \text{diag}(\bar{\boldsymbol{\Psi}}_b) \boldsymbol{\Theta}_b \mathbf{X}_{k^\prime,b} \right|^2.
\end{align}

The orientation optimization problem is formulated as
\begin{align}
	\text{(P1)}
	&\max_{ \mathbf{s}_b , \bar{\omega}_{b,q} , \mathbf{X}_{k,b} } \min_k \dot{g}_k( \theta_k^\text{e} , \phi_k^\text{e} ), \label{Problem 1} \\
	&\text{s.t.}\ \ \ \ \ \mathbf{s}_b[i] \in \{0,1\},\ \forall i, \tag{\ref{Problem 1}a}       \label{{Problem1}a} \\
	&\ \ \ \ \ \ \ \ \mathbf{1}^T\mathbf{s}_b = 1,\ \forall b, \tag{\ref{Problem 1}b}       \label{{Problem1}b}  \\
	&\ \ \ \ \ \ \ \ \sum_{q} \bar{\omega}_{b,q} = 1,\ \forall b, \tag{\ref{Problem 1}c}       \label{{Problem1}c} \\
	&\ \ \ \ \ \ \ \ \sum_b \Vert \mathbf{X}_b \Vert_2^2 \leq P_\text{tx},\ \forall b. \tag{\ref{Problem 1}d}       \label{{Problem1}d} \\
	&\ \ \ \ \ \ \ \ \eqref{eq7}-\eqref{eq9}.	\nonumber
\end{align}
The constraint \eqref{{Problem1}a} ensures that $\mathbf{s}_b$ is a binary vector, while the constraint \eqref{{Problem1}b} enforces that exactly one element of the vector $\mathbf{s}_b$ is equal to $1$, with all other elements being $0$. The constraint \eqref{{Problem1}c} ensures that the holographic beamformer is not greater than $1$ and the constraint \eqref{{Problem1}d} ensures that the transmit power of the 6DMHS-assisted transmitter is not greater than $P_\text{tx}$.The constraints \eqref{eq7}-\eqref{eq9} refer to  the rotation constraint for avoiding signal reflection, rotation constraint for avoiding signal blockage and minimum-distance constraint for avoiding collision, respectively. 
The optimization problem (P1) is nonconvex due to the nonconvex objective function and the binary constraint \eqref{{Problem1}a}. To address this problem, we adopt the alternating optimization algorithm to optimize $\mathbf{s}_b$ and adopt the FP-based algorithm to optimize $\bar{\omega}_{b,q}$ and $\mathbf{X}_{k,b}$. 

Firstly, by fixing $\bar{\omega}_{b,q}$ and $\mathbf{X}_{k,b}$, the optimization problem (P1) can be reformulated as
\begin{align}
	\text{(P2)}
	&\max_{ \mathbf{s}_b } \min_k \dot{g}_k( \theta_k^\text{e} , \phi_k^\text{e} ), \label{Problem 1} \\
	&\text{s.t.}\ \ \ \ \ \eqref{{Problem1}a},\ \eqref{{Problem1}b},\ \eqref{eq7}-\eqref{eq9}.	\nonumber
\end{align}
The optimization problem (P2) can be solved by using the alternating optimization algorithm \cite{9724245}, which is summarized as Algorithm \ref{alg:alg1}. The complexity for solving Algorithm \ref{alg:alg1} is $ \mathcal{O}\left( M_1 B K Q (M + K) \right) $.
\begin{algorithm}
	\linespread{1}\selectfont
	\caption{Alternating optimization algorithm for solving (P2).}
	\label{alg:alg1}
	\footnotesize
	\begin{algorithmic}[1]
		\REQUIRE\
		Initial values of $\mathbf{s}_b$, $\bar{\omega}_{b,q}$, $\mathbf{X}_{k,b}$;
		\ENSURE\
		The selection vector $\mathbf{s}_b$;
		\STATE Update $ t \leftarrow 1 $, $\overline{g}^\text{new} \leftarrow +\infty$, $ \overline{g}^\text{old} \leftarrow -\infty $, $\tilde{g}^\text{old} \leftarrow -\infty $, $\hat{\mathbf{s}}_b \leftarrow \mathbf{s}_b $;
		\WHILE{$ \left\vert \overline{g}^\text{new} - \overline{g}^\text{old} \right\vert \geq \epsilon $}
		\STATE Update $\overline{g}^\text{old} \leftarrow \overline{g}^\text{new} $;
		\FOR{$m_1=0$ to $M_1-1$ }
		\FOR{$b=0$ to $B-1$}
		\STATE Update $ \mathbf{s}_{b^\prime} \leftarrow \hat{\mathbf{s}}_{b^\prime} $, $\forall b^\prime=0,\cdots,B-1$;
		\STATE Update $\mathbf{s}_b \leftarrow \mathbf{0} $, $s_b[m_1] \leftarrow 1 $, $\tilde{g}^\text{new} \leftarrow \min_k \dot{g}_k(\theta_k^\text{e},\phi_k^\text{e}) $;
		\IF{$\tilde{g}^\text{new} > \tilde{g}^\text{old} $ and the constraints \eqref{eq7}-\eqref{eq9} are satisfied }
		\STATE Update $\tilde{g}^\text{old} \leftarrow \tilde{g}^\text{new} $, $\hat{\mathbf{s}}_b \leftarrow \mathbf{s}_b$;
		\ENDIF
		\ENDFOR
		\STATE Update $ \overline{g}^\text{new} \leftarrow \tilde{g}^\text{new} $;
		\ENDFOR
		\ENDWHILE
		\RETURN $\{ \mathbf{s}_b, \min_k \dot{g}_k(\theta_k^\text{e},\phi_k^\text{e}) \}$.
	\end{algorithmic}
\end{algorithm}

Secondly, by fixing $\mathbf{s}_b$, (P1) can be reformulated as
\begin{align}
	\text{(P3)}
	&\max_{  \bar{\omega}_{b,q} , \mathbf{X}_{k,b} } \min_k \dot{g}_k( \theta_k^\text{e} , \phi_k^\text{e} ), \label{Problem 3} \\
	&\text{s.t.}\ \ \ \ \ \eqref{{Problem1}c},\ \eqref{{Problem1}d}. \nonumber
\end{align}
By adopting the FP method, the objective function can be convert to be convex and (P3) can be solved effectively. Denote the vector $\boldsymbol{\Xi}_k \in \mathbb{C}^{K\times 1} $ as
	\begin{align}
		&\boldsymbol{\Xi}_k = \left[ \sum_b \mathbf{a}_b( \theta_{k}^\text{e} , \phi_{k}^\text{e}  ) \text{diag}(\bar{\boldsymbol{\Psi}}_b) \boldsymbol{\Theta}_b \mathbf{X}_{0,b} , \cdots , \right. \nonumber\\
		&\left.\ \ \ \ \ \ \  \sum_b \mathbf{a}_b( \theta_{k}^\text{e} , \phi_{k}^\text{e}  ) \text{diag}(\bar{\boldsymbol{\Psi}}_b) \boldsymbol{\Theta}_b \mathbf{X}_{K-1,b} \right]^T .
	\end{align}
where the vector  $\mathbf{a}_{b}(\theta,\phi)$ is given as
	\begin{align}
		&\mathbf{a}_{b}(\theta,\phi) = \sqrt{\frac{1}{M}} \cdot  \nonumber\\
		&   \left[ e^{ j \frac{2\pi }{\lambda} \mathbf{f}^T(\theta,\phi) ( \mathbf{s}_b^T\mathbf{q} + \mathbf{R}_b \bar{\mathbf{r}}_{0,0} ) }, \cdots, e^{ j \frac{2\pi }{\lambda} \mathbf{f}^T(\theta,\phi) ( \mathbf{s}_b^T\mathbf{q} + \mathbf{R}_b \bar{\mathbf{r}}_{0,M_y-1} ) } , \cdots, \right. \nonumber\\
		&  \left.  e^{ j \frac{2\pi }{\lambda} \mathbf{f}^T(\theta,\phi) ( \mathbf{s}_b^T\mathbf{q} + \mathbf{R}_b \bar{\mathbf{r}}_{M_x-1,0} ) } , \cdots , e^{ j \frac{2\pi }{\lambda} \mathbf{f}^T(\theta,\phi) ( \mathbf{s}_b^T\mathbf{q} + \mathbf{R}_b \bar{\mathbf{r}}_{M_x-1,M_y-1} ) } \right]^T,
	\end{align}
Then, by introducing the auxiliary variable $\boldsymbol{\zeta}_k\in\mathbb{C}^{K\times1}$, the objective function of (P3) can be reformulated as
\begin{align}
	\ddot{g}_k(\theta_k^\text{e},\phi_k^\text{e}) = 2 \Re\left\{ \boldsymbol{\zeta}_k^H \boldsymbol{\Xi}_k\right\} - \boldsymbol{\zeta}_k^H \boldsymbol{\zeta}_k.
\end{align}
The problem (P3) can be reformulated as 
\begin{align}
	\text{(P4)}
	&\max_{  \bar{\omega}_{b,q} , \mathbf{X}_{k,b} , \boldsymbol{\zeta}_k } \min_k \ddot{g}_k( \theta_k^\text{e} , \phi_k^\text{e} ), \label{Problem 4} \\
	&\text{s.t.}\ \ \ \ \ \eqref{{Problem1}c},\ \eqref{{Problem1}d}. \nonumber
\end{align}
Next, we solve (P4) by alternatively optimizing $\bar{\omega}_{b,q}$, $\mathbf{X}_{k,b}$ and $\boldsymbol{\zeta}_k$. 

By fixing $\bar{\omega}_{b,q}$ and $\boldsymbol{\zeta}_k$, (P4) can be reformulated as
\begin{align}
	\text{(P4-1)}
	&\max_{   \mathbf{X}_{k,b} } \min_k \ddot{g}_k ( \theta_k^\text{e} , \phi_k^\text{e} ), \label{Problem 3} \\
	&\text{s.t.}\ \ \ \ \ \eqref{{Problem1}d}. \nonumber
\end{align}

By fixing $\mathbf{X}_{k,b}$ and $\boldsymbol{\zeta}_k$, (P4) can be reformulated as
\begin{align}
	\text{(P4-2)}
	&\max_{   \bar{\omega}_{b,q} } \min_k \ddot{g}_k( \theta_k^\text{e} , \phi_k^\text{e} ), \label{Problem 3} \\
	&\text{s.t.}\ \ \ \ \ \eqref{{Problem1}c}. \nonumber
\end{align}
The problems (P4-1) and (P4-2) are convex and can be solved by adopting the interior point method \cite{Nocedal2006}. 

By fixing $\bar{\omega}_{b,q}$ and $\mathbf{X}_{k,b}$, (P4) can be reformulated as
\begin{align}
	\text{(P4-3)}
	&\max_{ \boldsymbol{\zeta}_k } \min_k \ddot{g}_k( \theta_k^\text{e} , \phi_k^\text{e} ). \label{Problem 3} 
\end{align}
By adopting the method of derivation, the optimal $\boldsymbol{\zeta}_k$ is given as
\begin{align}
	\boldsymbol{\zeta}_k = \boldsymbol{\Xi}_k. \label{eq53}
\end{align}

By alternatively optimizing $\bar{\omega}_{b,q}$, $\mathbf{X}_{k,b}$ and $\boldsymbol{\Xi}_k$, the optimal solution to (P4) can be obtained, which also yields  the optimal value of (P3). Then, by alternatively solving (P2) and (P3), the optimal solution of (P1) is obtained.
The algorithm for solving (P1) is summarized in Algorithm \ref{alg:alg2}.
The complexity of Algorithm \ref{alg:alg2} is $\mathcal{O}\left( T_{\text{ite},1}\left( M_1BKQ(M+K) + T_{\text{ite},2} B^{3.5} K^{3.5} Q^{3.5} \right)  \right)$, where $T_{\text{ite},1}$ and $T_{\text{ite},2}$ represent the number of the inner iterations and the number of the outer iterations, respectively.

\begin{algorithm}
	\linespread{1}\selectfont
	\caption{ Orientation optimization algorithm.}
	\label{alg:alg2}
	\footnotesize
	\begin{algorithmic}[1]
		\REQUIRE\
		Initial values of $\mathbf{s}_b$, $\bar{\omega}_{b,q}$, $\mathbf{X}_{b,k}$, $\boldsymbol{\zeta}_k$;
		\ENSURE\
		The selection vector $\mathbf{s}_b$ and the weight factor $\bar{\omega}_{b,q}$;
		\STATE Update $\overline{g}^\text{new} \leftarrow +\infty$, $ \overline{g}^\text{old} \leftarrow -\infty $, $\tilde{g}^\text{new} \leftarrow +\infty $, $\tilde{g}^\text{old} \leftarrow -\infty $;
		\WHILE{$ \left\vert \overline{g}^\text{new} - \overline{g}^\text{old} \right\vert \geq \epsilon $}
		\STATE Update $ \overline{g}^\text{old} \leftarrow \overline{g}^\text{new} $;
		\STATE Update $\mathbf{s}_b$ by adopting Algorithm 1;
		\WHILE{$ \left\vert \tilde{g}^\text{new} - \tilde{g}^\text{old} \right\vert \geq \epsilon $}
		\STATE Update $\tilde{g}^\text{old} \leftarrow \tilde{g}^\text{new}$;
		\STATE Update $\mathbf{X}_{k,b}$ by solving (P4-1);
		\STATE Update $\omega_{b,q}$ by solving (P4-2);
		\STATE Update $\boldsymbol{\zeta}_k$ by Eq. \eqref{eq53};
		\STATE Update $\tilde{g}^\text{new} \leftarrow \min_k \ddot{g}_k(\theta_k^\text{e},\phi_k^\text{e})$;
		\ENDWHILE
		\STATE Update $\overline{g}^\text{new} \leftarrow \min_k \dot{g}_k( \theta_k^\text{e} , \phi_k^\text{e} ) $;
		\ENDWHILE
		\RETURN $\{ \mathbf{s}_b,\ \bar{\omega}_{b,q},\  \mathbf{X}_{b,k},\ \boldsymbol{\zeta}_k,\ \min_k \dot{g}_k(\theta_k^\text{e} , \phi_k^\text{e} ) \}$.
	\end{algorithmic}
\end{algorithm}

\section{Joint Beamforming and Power Splitter Design}
In this section, we jointly optimize the digital beamforming and the power splitter based on the equivalent CSI. During the orientation adjustment stage, the optimal orientation of the 6DMHS and the holographic beamforming of the RHS are determined. Subsequently, in the downlink transmission stage, the equivalent CSI at each feed is estimated and exploited for IDET services. According to \eqref{eq10}, the equivalent channel $\bar{\mathbf{h}}_{k,b} \in \mathbb{C}^{1\times Q}$ is expressed as
\begin{align}
	\bar{\mathbf{h}}_{k,b} 
	= \sqrt{M} \sum_\iota \Lambda_{k,\iota,b} \eta_{k,\iota,b} 
	\mathbf{a}^T_{b}(\theta_{k,\iota,b},\phi_{k,\iota,b}) 
	\mathrm{diag}( \boldsymbol{\Psi}_b ) \boldsymbol{\Theta}_b.
\end{align}
In the sequel, the equivalent CSI is estimated within each short IDET frame, after which the downlink digital beamforming and power splitter are jointly designed to maximize the IDET performance.

The minimal EH power maximization problem is formulated as
\begin{align}
	\text{(P5)}
	&\max_{ \mathbf{X}_{k,b}, \rho_k } \min_k \Gamma(P_{\text{EH},k}), \label{Problem 5} \\
	&\text{s.t.}\ \ \ \ \ R_k \geq R_0, \tag{\ref{Problem 5}a}       \label{{Problem5}a} \\
	&\ \ \ \ \ \ \ \ \sum_{b,k} \left\Vert \mathbf{X}_{k,b} \right\Vert_2^2 \leq P_\text{tx}.\tag{\ref{Problem 5}b}       \label{{Problem5}b} \\
	&\ \ \ \ \ \ \ \ 0 \leq \rho_k \leq 1, \tag{\ref{Problem 5}c} \label{{Problem5}c} 
\end{align}
(P5) is a nonconvex problem, due to the objective function and the constraint \eqref{{Problem5}b} are nonconvex. 

Firstly, by introducing the auxiliary variable ${\vartheta}_k \in \mathbb{C}^{1} $, $\gamma_k$ can be reformulated as
\begin{small}
	\begin{align}
		\dot{\gamma}_k = 
		& 2 \Re\left\{ \sqrt{1-\rho_k} \vartheta_k^H \sum_b \overline{ \mathbf{h} }_{k,b} \mathbf{X}_{k,b} \right\} \nonumber \\
		& - \vartheta_k^H\vartheta_k \left\{ (1-\rho_k) \sum_{k^\prime \neq k} \left| \sum_b \overline{\mathbf{h}}_{k,b} \mathbf{X}_{k^\prime,b} \right|^2 + (1-\rho_k) \sigma_0^2 + \sigma_\text{cov}^2   \right\}.
	\end{align}
\end{small}
The downlink throughput $R_k$ can be reformulated as 
\begin{align}
	\dot{R}_k = \log_2\left( 1 + \dot{\gamma}_k \right),\ \text{[bit/s/Hz]}.
\end{align}

Secondly, by introducing the variable  $\boldsymbol{\varsigma}_k \in \mathbb{C}^{K\times1}$, $P_{\text{EH},k}$ is reformulated as
\begin{align}
	\dot{P}_{\text{EH},k} = 2 \Re\{ \boldsymbol{\varsigma}_k^H \dot{\boldsymbol{\Xi}}_k \} - \boldsymbol{\varsigma}_k^H \boldsymbol{\varsigma}_k + \rho_k \sigma_0^2,
\end{align}
where $\dot{\boldsymbol{\Xi}}_k  = \sqrt{\rho_k} \left[ \sum_b \overline{\mathbf{h}}_{k,b} \mathbf{X}_{b,0} , \cdots , \sum_b \overline{\mathbf{h}}_{k,b} \mathbf{X}_{b,K-1} \right]^T \in \mathbb{C}^{K\times1}$. 

Then, (P5) can be reformulated as
\begin{align}
	\text{(P6)}
	&\max_{ \mathbf{X}_{k,b}, \rho_k , \vartheta_k , {\mathbf{\varsigma}}_k } P_0, \label{Problem 6} \\
	&\text{s.t.}\ \ \ \ \  
	\Gamma(\dot{P}_{\text{EH},k}) \geq P_0 \tag{\ref{Problem 6}a}       \label{{Problem6}a}\\
	&\ \ \ \ \ \ \ \
	\dot{R}_k \geq R_0, \tag{\ref{Problem 6}b}       \label{{Problem6}b} \\
	&\ \ \ \ \ \ \ \ \eqref{{Problem5}b}-\eqref{{Problem5}c}. \nonumber
\end{align}
The constraint \eqref{{Problem6}a} is nonconvex. By using the inverse function of the function $\Gamma(\dot{P}_{\text{EH},k})$, the constraint \eqref{{Problem6}a} can be reformulated as
\begin{align}
	\dot{P}_{\text{EH},k} \geq \Gamma^{-1}(P_0),
\end{align}
where $\Gamma^{-1}(\cdot)$ is the inverse function of $\Gamma(\cdot)$. By letting $\dot{P}_0 = \Gamma^{-1}(P_0)$, (P6) can be reformulated as
\begin{align}
	\text{(P7)}
	&\max_{ \mathbf{X}_{k,b}, \rho_k , \vartheta_k , {\mathbf{\varsigma}}_k } \dot{P}_0, \label{Problem 7} \\
	&\text{s.t.}\ \ \ \ \  
	\dot{P}_{\text{EH},k} \geq \dot{P}_0 \tag{\ref{Problem 7}a}       \label{{Problem7}a}\\
	&\ \ \ \ \ \ \ \ \eqref{{Problem6}b},\ \eqref{{Problem5}b}-\eqref{{Problem5}c}. \nonumber
\end{align}
Next, by alternatively optimizing $\mathbf{X}_{k,b}, \rho_k , \vartheta_k$ and  ${\mathbf{\varsigma}}_k$, (P7) can be solved effectively. 

By fixing $ \rho_k , \vartheta_k$ and  ${\mathbf{\varsigma}}_k$, (P7) is reformulated as
\begin{align}
	\text{(P7-1)}
	&\max_{ \mathbf{X}_{k,b} } \dot{P}_0, \label{Problem 7-1} \\
	&\text{s.t.}\ \ \ \ \  \eqref{{Problem7}a},\ \eqref{{Problem6}b}, \ \eqref{{Problem5}b}.
	\nonumber
\end{align}
By fixing $\mathbf{X}_{k,b}, \vartheta_k$ and  ${\mathbf{\varsigma}}_k$, (P7) can be reformulated as
\begin{align}
	\text{(P7-2)}
	&\max_{ \rho_k } \dot{P}_0, \label{Problem 7-2} \\
	&\text{s.t.}\ \ \ \ \  \eqref{{Problem7}a},\ \eqref{{Problem6}b}, \ \eqref{{Problem5}c}.
	\nonumber
\end{align}
The problems (P7-1) and (P7-2) are convex, which can be solved by using the interior point method. By fixing $\mathbf{X}_{k,b}, \rho_k$ and  ${\mathbf{\varsigma}}_k$, (P7) can be reformulated as
\begin{align}
	\text{(P7-3)}
	&\max_{ \vartheta_k } \dot{P}_0, \label{Problem 7-3} \\
	&\text{s.t.}\ \ \ \ \  \eqref{{Problem6}b}.
	\nonumber
\end{align}
By adopting the method of derivation, the optimal $\vartheta^*_k$ is given as
\begin{align}
	\vartheta^*_k = \frac{\sqrt{1-\rho_k} \sum_b \overline{ \mathbf{h} }_{k,b} \mathbf{X}_{k,b} }{ (1-\rho_k) \sum_{k^\prime \neq k} \left| \sum_b \overline{\mathbf{h}}_{k,b} \mathbf{X}_{k^\prime,b} \right|^2 + (1-\rho_k) \sigma_0^2 + \sigma_\text{cov}^2 }. \label{eq59}
\end{align}
By fixing $\mathbf{X}_{k,b}, \rho_k$ and $ \vartheta_k$, (P7) can be reformulated as
\begin{align}
	\text{(P7-4)}
	&\max_{ {\mathbf{\varsigma}}_k } \dot{P}_0, \label{Problem 7-4} \\
	&\text{s.t.}\ \ \ \ \  \eqref{{Problem7}a}.
	\nonumber
\end{align}
Similarly, by adopting the method of derivation, the optimal ${\mathbf{\varsigma}}_k^*$ is given as
\begin{align}
	{\mathbf{\varsigma}}_k^* = \dot{\mathbf{\Xi}}_k. \label{eq61}
\end{align}
By alternatively solving (P7-1)-(P7-4), the optimal solutions of (P7) are obtained, which are also the optimal solutions of (P5) and (P6). The FP-based algorithm for solving (P5) is summarized in Algorithm \ref{alg:alg3}. The complexity of the Algorithm \ref{alg:alg3} is $\mathcal{O}\left( T_{\text{ite},3}\left(  B^{3.5}Q^{3.5}K^{3.5} \right)  \right)$, where $T_{\text{ite},3}$ represents the number of the iterations.

\begin{algorithm}
	\linespread{1}\selectfont
	\caption{ FP-based optimization algorithm for solving (P2).}
	\label{alg:alg3}
	\footnotesize
	\begin{algorithmic}[1]
		\REQUIRE\
		Initial value of $\mathbf{X}_{k,b}$, $\rho_k$, $\vartheta_k$ and  $\boldsymbol{\varsigma}_k$;
		\ENSURE\
		The digital beamforming vector $\mathbf{X}_{k,b}$ and the power splitting factor $\rho_k$;
		\STATE Update $\overline{P}_{0}^\text{new} \leftarrow +\infty$, $ \overline{P}_0^\text{old} \leftarrow -\infty $;
		\WHILE{$ \left\vert \overline{P}_0^\text{new} - \overline{P}_0^\text{old} \right\vert \geq \epsilon $}
		\STATE Update $ \overline{P}_0^\text{old} \leftarrow \overline{P}_0^\text{new} $;
		\STATE Update $\mathbf{X}_{k,b}$ by solving (P7-1);
		\STATE Update $\rho_k$ by solving (P7-2);
		\STATE Update $\vartheta_k$ by Eq. \eqref{eq59};
		\STATE Update $\boldsymbol{\varsigma}_k$ by Eq. \eqref{eq61};
		\STATE Update $\overline{P}_0^\text{new} \leftarrow \min_k \Gamma(P_{\text{EH},k}) $;
		\ENDWHILE
		\RETURN $\left\{ \mathbf{X}_{k,b},\ \rho_k,\ \vartheta_k,\ \boldsymbol{\varsigma}_k,\ \min_k \Gamma(P_{\text{EH},k}) \right\}$.
	\end{algorithmic}
\end{algorithm}

\section{Simulation Results}
In this section, numerical results are presented to validate the performance of the proposed 6DMHS-assisted IDET system. The 3GPP channel model is employed, wherein the channel between the 6DMHS-assisted transmitter and the IDET receivers comprises one LoS path and three NLoS paths. The parameters are set as follows: $f_c = 30 $ GHz, $c = 3 \times 10^8$ m/s, $ M = 32 \times 32 $, $Q = 1$, $M_1 = 50$, $P_t = 40$ dBm, $K_\text{R} = 10$, $\sigma_0^2 = -100$ dBm and $ \sigma_\text{cov}^2 = -50$ dBm. 

Some benchmarks are listed as follows:
\begin{itemize}
    \item \textbf{FPA:} In this scheme, the positions and rotations of all the RHS are fixed. The RHS is uniformly distributed on the surface of a sphere with a radius of 1 m. Furthermore, the orientation of each RHS is aligned with the normal vector pointing from the BS to the center of the corresponding RHS.
    \item \textbf{6DMHS with rotation only:} In this scheme, the positions of RHS is fixed and each RHS is uniformly placed on the surface of a sphere with a radius of 1 m. Additionally, the orientation of each RHS is aligned with the sensing direction of its corresponding IDET receiver.
    \item \textbf{6DMHS with translation only:} In this scheme, the rotations of RHS is fixed along the normal vector direction from the BS to the center of each RHS. The discrete center positions of the RHS are located on the surface of a sphere with a radius of 1 m and are optimized to maximize the beamforming gain.
    \item \textbf{Least square (LS)-based sensing  \cite{10097166}:} This scheme utilizes the least square-based sensing method to estimate the angles of the IDET receivers, while the positions and rotations of the 6DMHS-assisted transmitter are optimized using Algorithm \ref{alg:alg2} proposed in this paper.
\end{itemize}

\begin{figure}
	\centering
	\includegraphics[width=1\linewidth]{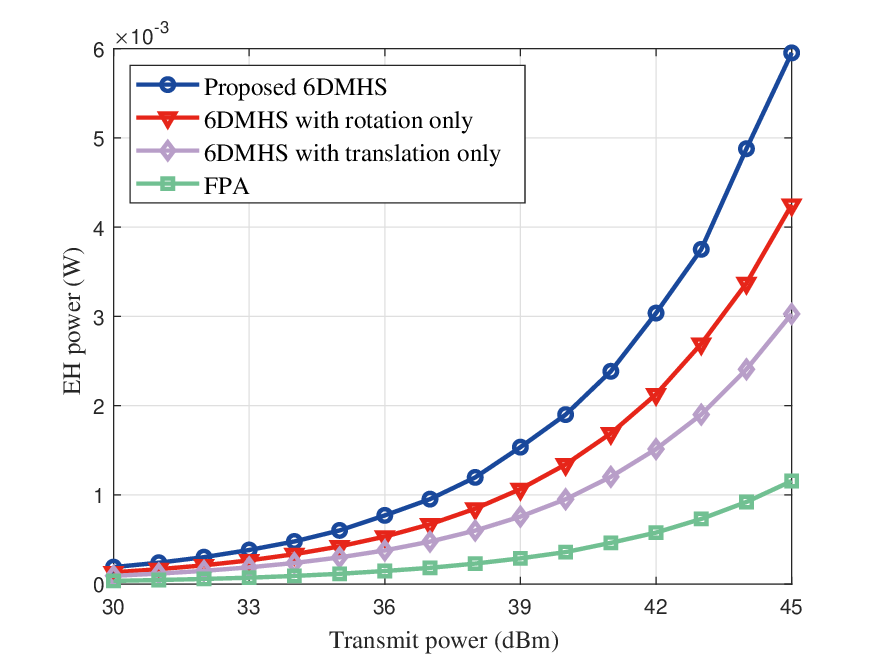}
	\setlength{\abovecaptionskip}{0pt}
	\setlength{\belowcaptionskip}{0pt} \caption{ EH power performance versus the transmit power under different transmitter configurations.} \label{fig4}
\end{figure}

Fig. \ref{fig4} illustrates the variation of the EH power with respect to the downlink transmit power. As the transmit power increases, the EH power gradually improves. Meanwhile, the proposed scheme is compared with the \textbf{6DMHS with rotation only}, \textbf{6DMHS with translation only} and \textbf{FPA} benchmarks. The proposed scheme achieves enhanced spatial degrees of freedom by jointly optimizing the rotations and translations of the 6DMHS, thereby attaining the best IDET performance. The \textbf{6DMHS with rotation only} scheme outperforms the \textbf{6DMHS with translation only} scheme, as the rotation operation enables alignment of the direction of maximum beamforming gain with the corresponding IDET receiver, resulting in improved IDET performance. Furthermore, both the \textbf{6DMHS with rotation only} and \textbf{6DMHS with translation only} schemes outperform the \textbf{FPA} scheme, since the added mobility in these schemes provide greater flexibility, leading to superior IDET performance.

\begin{figure}
	\centering
	\includegraphics[width=1\linewidth]{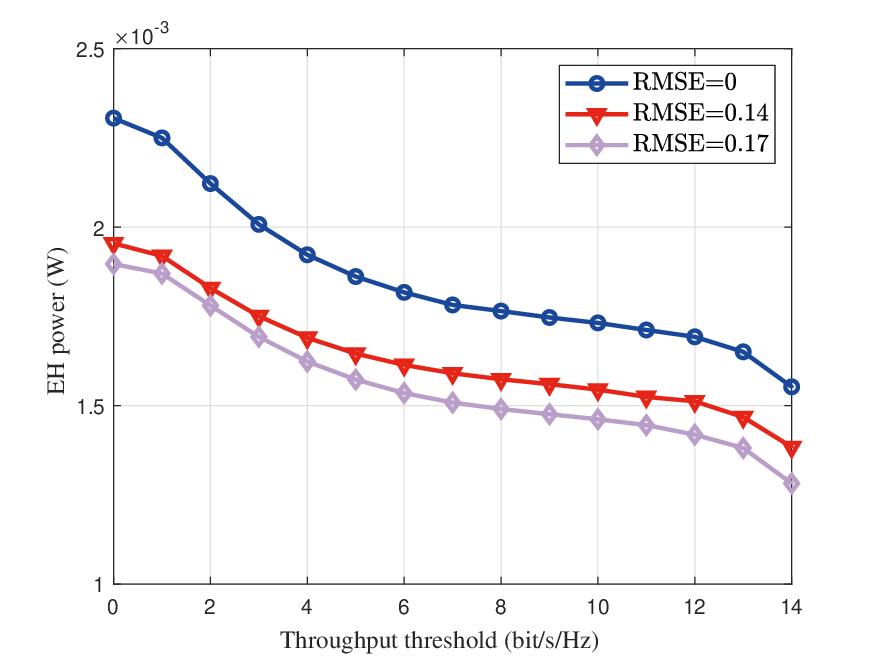}
	\setlength{\abovecaptionskip}{0pt}
	\setlength{\belowcaptionskip}{0pt} \caption{ EH power performance versus the throughput threshold under different sensing errors.} \label{fig5}
\end{figure}

Fig. \ref{fig5} illustrates the trade-off between the downlink throughput threshold and the EH power. 
Firstly, we define the root-mean-square error of the sensing angles as $\text{RMSE} = \frac{1}{T_\text{S}} \sum_{t_\text{S}=0}^{T_\text{S}-1} \left( \| \mathbf{f}(\theta_k^{\text{e}} , \phi_k^{\text{e}} ) - \mathbf{f}(\theta_k^{\text{t}} , \phi_k^{\text{t}} ) \|_2^2 \right) $, where $\theta_k^{\text{t}}$ and $\phi_k^{\text{t}}$ are the true azimuth and elevation angles of the $k$-th IDET receiver or its scatterer with the maximal channel power gain.
As the throughput threshold increases, the EH power gradually decreases. This is because a larger portion of the received RF power for the IDET receiver is allocated for information decoding, thereby reducing the input power of the energy harvester. Additionally, the impact of angular sensing accuracy on the IDET performance is analyzed. When there is no sensing error, \textit{i.e.}, $\text{RMSE} = 0$, the optimal IDET performance is achieved. As the sensing error increases from $\text{RMSE} = 0$ to $\text{RMSE} = 0.17$, the performance steadily degrades. This result highlights the critical role of uplink sensing accuracy in downlink throughput, indicating that the higher sensing precision leads to an improved IDET performance.

\begin{figure}
	\centering
	\includegraphics[width=1\linewidth]{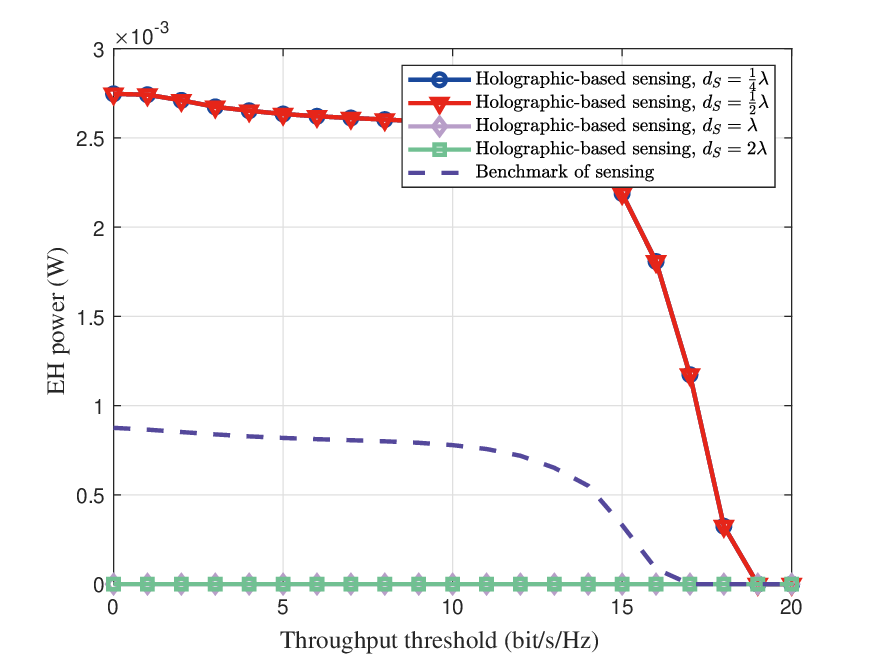}
	\setlength{\abovecaptionskip}{0pt}
	\setlength{\belowcaptionskip}{0pt} \caption{ EH power performance versus the adjustment factor under sensing methods.} \label{fig6}
\end{figure}

\begin{figure}
	\centering
	\includegraphics[width=1\linewidth]{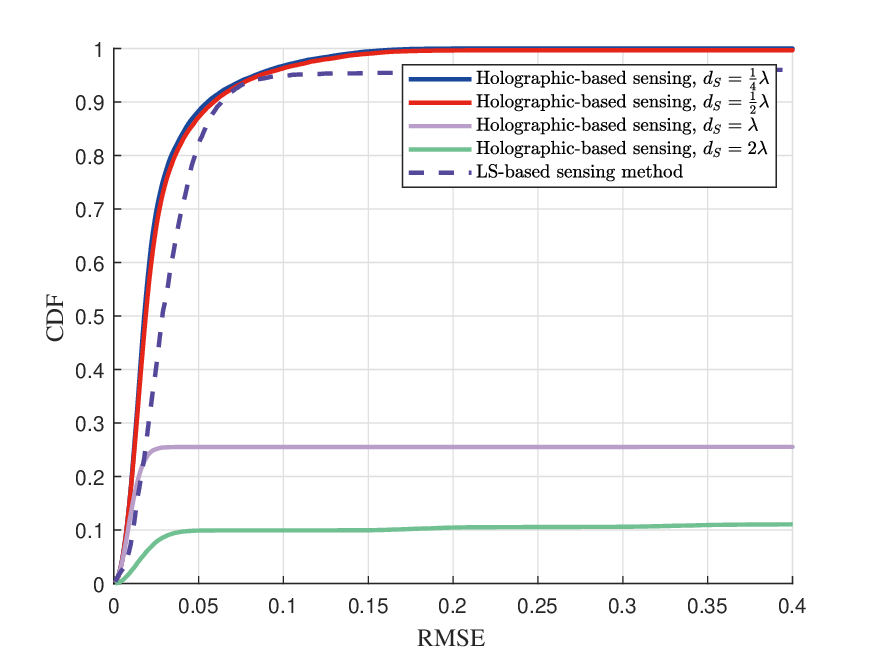}
	\setlength{\abovecaptionskip}{0pt}
	\setlength{\belowcaptionskip}{0pt} \caption{CDF of the sensing error under different $\kappa$ and different sensing methods.} \label{fig7}
\end{figure}

Fig. \ref{fig6} illustrates the impact of the throughput threshold on the IDET performance under different sensing elements spacing and different sensing scheme. It can be observed that when adopts the proposed sensing method and $d_\text{S}\leq \frac{\lambda}{2}$, the IDET performance is approximately equivalent under different spacing of sensing elements. Moreover, the proposed holographic-based sensing method consistently outperforms the LS-based sensing method in terms of IDET performance, indicating that our proposed sensing method has a better sensing performance. However, when $d_\text{S}>\frac{\lambda}{2}$, the IDET performance declines sharply. This degradation is attributed to the deterioration in sensing accuracy, as detailed below.
Fig. \ref{fig7} presents the sensing accuracy under different sensing element spacing. When $d_\text{S}\leq\frac{\lambda}{2}$, the proposed holographic-based sensing method achieves at least a $95\%$ probability that the RMSE remains below $0.1$. In contrast, when $d_\text{S}>\frac{\lambda}{2}$, the sensing accuracy of the proposed method deteriorates significantly. This observation supports Lemma 2, which states that placing sensing elements at half-wavelength intervals is sufficient to achieve near-optimal sensing performance.

Additionally, Fig. \ref{fig7} shows that the \textbf{LS-based sensing } method performs worse than the proposed holographic-based sensing method. This is because the \textbf{LS-based sensing} method derives sensing information from the signals of the feeds, which aggregates the responses of all RHS elements. The signal aliasing resulting from this aggregation degrades angular estimation accuracy. Moreover, the lacks of the imaginary part of the holographic beamformer may hurt the sensing performance of the \textbf{LS-based sensing } method. Consequently, the IDET performance adopting the \textbf{LS-based sensing} method reduces. In contrast, the proposed sensing method acquires sensing data in parallel from individual sensing elements, thereby avoiding inter-element aliasing and achieving superior sensing accuracy. As a result, the proposed holographic-based sensing method enables a higher IDET performance.

\begin{figure}
	\centering
	\includegraphics[width=1\linewidth]{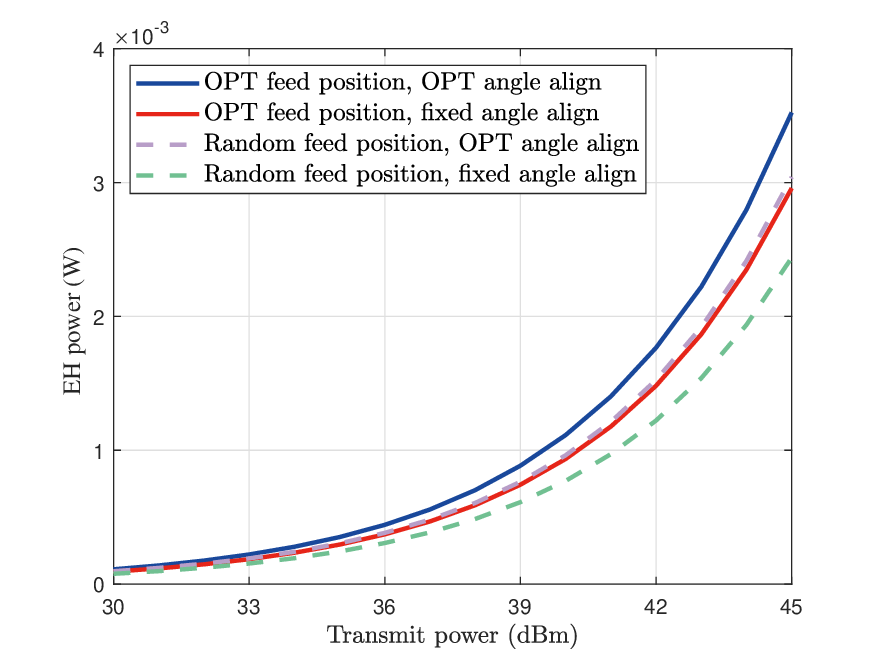}
	\setlength{\abovecaptionskip}{0pt}
	\setlength{\belowcaptionskip}{0pt} \caption{EH power versus the transmit power under different angle alignment and feed positioning methods.} \label{fig8}
\end{figure}

Fig. \ref{fig8} compares the IDET performance under different feed position configurations and alignment strategies. It is observed that aligning the maximum beamforming gain direction of the RHS with the IDET receivers yields better IDET performance than simply aligning the normal vector of the RHS with the receiver direction. This confirms that directing the RHS's peak beamforming gain toward the receivers can significantly enhance the IDET performance, thereby validating the effectiveness of the proposed alignment strategy.  
Moreover, optimizing the feed positions further improves the IDET performance. This is because the optimized feed positions enable the RHS elements to achieve coherent signal combination, thereby increasing the radiated power in the direction of the main beam. As a result, the IDET performance is substantially enhanced. These results highlight the importance of feed position optimization during the RHS design and preparation process.

\begin{figure}
	\centering
	\includegraphics[width=1\linewidth]{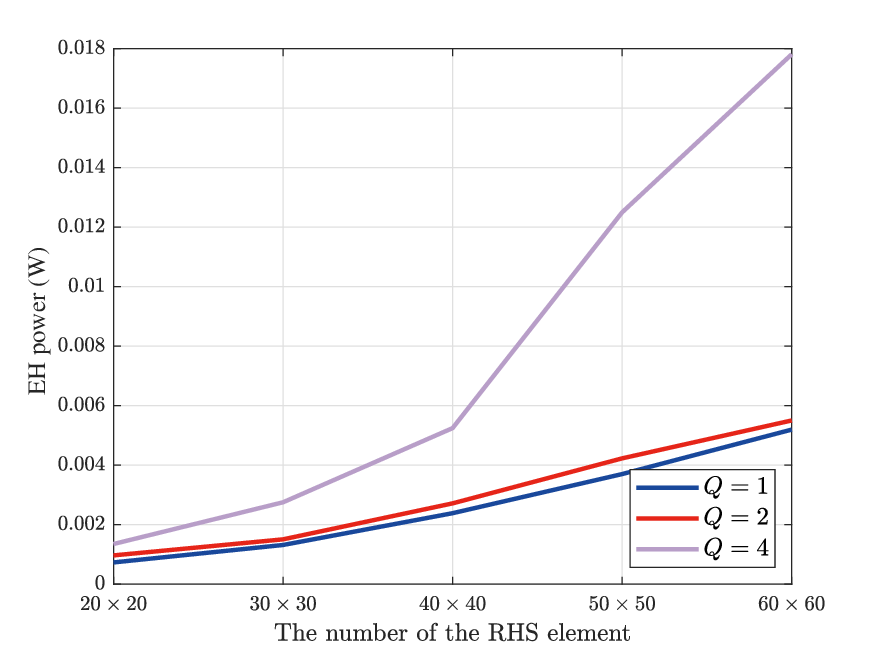}
	\setlength{\abovecaptionskip}{0pt}
	\setlength{\belowcaptionskip}{0pt} \caption{The EH power performance versus the number of the RHS element under different number of feeds.} \label{fig9}
\end{figure}

Fig. \ref{fig9} illustrates the impact of the number of the RHS elements and feeds on the IDET performance. It is observed that as the number of elements increases, the EH power increases monotonically. This is because more elements contribute to a higher spatial gain, thereby enhancing the IDET performance. Moreover, increasing the number of feeds from $1$ to $4$ leads to a gradual improvement in IDET performance, due to the greater degrees of freedom in digital beamforming.

\begin{figure}
\centering
\includegraphics[width=1\linewidth]{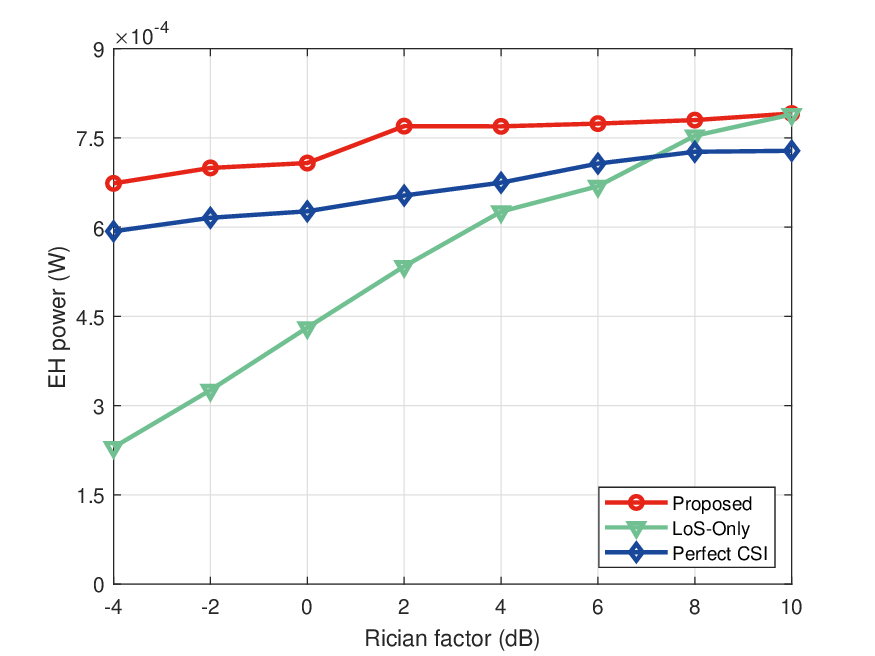}
\setlength{\abovecaptionskip}{0pt}
\setlength{\belowcaptionskip}{0pt} \caption{EH power performance versus the rice factor under different protocols.} \label{fig10}
\end{figure}

Assume that the CSI remains unchanged within a coherence time block of $T_c = 120$. For our proposed scheme, the pilot overhead is given by $T_{s,1} = \alpha K S \log_2(BQ)$, and remaining time allocated for IDET transmission is $T_c - T_{s,1}$, where $\alpha=0.32$ is a constant parameter \cite{9400843}. In contrast, for the perfect CSI scheme, the pilot overhead is $ T_{s,2} = \alpha K S \log_2(BM) $ \cite{9400843} and the transmission time is $T_c-T_{s,2}$. Since $Q\ll M$, it follows that $T_{s,1} \ll T_{s,2}$. Fig. \ref{fig10} illustrates the impact of the Rician factor on IDET performance. 
As the Rician factor increases from $-4$ dB to $10$ dB, the IDET performance of the \textbf{Proposed} scheme improves monotonically. This is because a larger Rician factor gradually concentrates the channel power gain on the LoS component, thereby enhancing the IDET performance.
Additionally, we compare the \textbf{Proposed} scheme with the \textbf{LoS-Only} scheme. It is observed that when the Rician factor is $10$ dB, the \textbf{LoS-Only} scheme achieves nearly the same IDET performance as the proposed scheme. This is because, at a very high Rician factor, the LoS link becomes the dominant component of the channel and the NLoS components can be ignored, leading to comparable IDET performance. However, when the Rician factor varies from $-4$ dB to $8$ dB, the IDET performance of the \textbf{LoS-Only} scheme is inferior to that of the \textbf{Proposed} scheme. This is because the \textbf{Proposed} scheme designs the holographic beamformer by exploiting the NLoS link with the maximal channel power gain, which can exceed the channel power gain of the LoS link.
Moreover, considering the overhead of channel estimation, the \textbf{Proposed} scheme outperforms the \textbf{Perfect CSI} scheme. This is because the channel power gain is concentrated either on the LoS link or on the NLoS link with the maximal channel power gain, and thus designing the holographic beamformer with perfect CSI may not significantly improve the IDET performance. Meanwhile, the substantial overhead required by the \textbf{Perfect CSI} scheme reduces its overall performance. It is also observed that when the Rician factor is below $7$ dB, the \textbf{Perfect CSI} scheme outperforms the \textbf{LoS-Only} scheme, since at low Rician factors the NLoS components degrade the IDET performance and carefully accounting for them can improve the received signal quality. In contrast, when the Rician factor exceeds $7$ dB, the LoS link becomes dominant, and excessive CSI estimation may even hurt the IDET performance, thereby making the \textbf{LoS-Only} scheme superior to the \textbf{Perfect CSI} scheme.

%As the Rician factor decreases, the IDET performance declines monotonically. This is because a lower Rician factor implies that the channel energy shifts from the LoS path to NLoS paths, introducing more multi-path interference that degrades the IDET performance. Additionally, the IDET performance achieved with perfect CSI is inferior to that of the proposed scheme. As previously discussed, this is mainly due to the substantial pilot overhead associated with the perfect CSI acquisition, which reduces the effective transmission time and thus harms the IDET performance. These results demonstrate that the proposed design can significantly reduces the pilot overhead and effectively improving the IDET performance.

\section{Conclusion}
This paper investigated a sensing-enhanced 6DMHS-assisted IDET system. A holographic-based sensing method was proposed to acquire the angular information of IDET receivers. Leveraging the obtained sensing information, the holographic beamforming of the RHS together with the rotations and positions of the 6DMHS were jointly optimized to align the maximum beamforming direction of the RHS with the IDET receiver. Subsequently, by fixing the holographic beamforming and the 6DMHS orientation, the downlink equivalent CSI was obtained, while the digital beamforming and power splitting factor were further optimized. As a result, the IDET performance was significantly improved. Simulation results show that: 1) The proposed parallel-form holographic sensing method achieves superior sensing performance over the conventional series-form method; 2) Aligning the maximum beamforming direction of the RHS with the IDET receiver yields notable IDET performance gains; 3) The proposed three-stage protocol outperforms the traditional perfect-CSI benchmark in the presence of an absolutely dominant path in the wireless channel, owing to its reduced pilot overhead.

\begin{appendices}
\label{AppendixA}
	\section{Proof of Theorem 1}
	Firstly, the limitation  of $\lim_{N\rightarrow\infty}\frac{1}{N} \sum_{n_x,n_y} e^{ j \frac{2\pi}{\lambda} \mathbf{f}^T(\theta_{k,b}^\text{L,e} ,\phi_{k,b}^\text{L,e} ) \mathbf{r}_{b,n_x,n_y}^\text{S,L}  } h_{k,b,n_x,n_y}^\text{S,L}$ is expressed as Eq. \eqref{eq71}, where $\mathcal{X}_1$, $\mathcal{X}_2$, $\mathcal{X}_3$ and $\mathcal{X}_4$ are given as
	\begin{align}
	\mathcal{X}_1  =&   \frac{-N_x+1}{2}d_\text{S} \left( -f_1(\theta_{k,b}^\text{L,e},\phi_{k,b}^\text{L,e}) + f_1(\theta_{k,\iota,b}^\text{L},\phi_{k,\iota,b}^\text{L}) \right) \nonumber\\
	&+ \frac{-N_y+1}{2}d_\text{S} \left( -f_2(\theta_{k,b}^\text{L,e},\phi_{k,b}^\text{L,e}) + f_2(\theta_{k,\iota,b}^\text{L},\phi_{k,\iota,b}^\text{L}) \right) ,\\
	\mathcal{X}_2  =&   \frac{-N_x+1}{2}d_\text{S} \left( -f_1(\theta_{k,b}^\text{L,e},\phi_{k,b}^\text{L,e}) + f_1(\theta_{k,\iota,b}^\text{L},\phi_{k,\iota,b}^\text{L}) \right) \nonumber\\
	&+ \frac{N_y-1}{2}d_\text{S} \left( -f_2(\theta_{k,b}^\text{L,e},\phi_{k,b}^\text{L,e}) + f_2(\theta_{k,\iota,b}^\text{L},\phi_{k,\iota,b}^\text{L}) \right),\\
	\mathcal{X}_3  =&   \frac{N_x-1}{2}d_\text{S} \left( -f_1(\theta_{k,b}^\text{L,e},\phi_{k,b}^\text{L,e}) + f_1(\theta_{k,\iota,b}^\text{L},\phi_{k,\iota,b}^\text{L}) \right) \nonumber\\
	& +  \frac{-N_y+1}{2}d_\text{S} \left( -f_2(\theta_{k,b}^\text{L,e},\phi_{k,b}^\text{L,e}) + f_2(\theta_{k,\iota,b}^\text{L},\phi_{k,\iota,b}^\text{L}) \right) ,\\
	\mathcal{X}_4  =  & \frac{N_x-1}{2}d_\text{S} \left( -f_1(\theta_{k,b}^\text{L,e},\phi_{k,b}^\text{L,e}) + f_1(\theta_{k,\iota,b}^\text{L},\phi_{k,\iota,b}^\text{L}) \right) \nonumber\\
	& + \frac{N_y-1}{2}d_\text{S} \left( -f_2(\theta_{k,b}^\text{L,e},\phi_{k,b}^\text{L,e}) + f_2(\theta_{k,\iota,b}^\text{L},\phi_{k,\iota,b}^\text{L}) \right) .
\end{align}

	\begin{figure*}[ht!]
		\begin{align}
			&\frac{1}{N} \sum_{n_x,n_y} e^{ - j \frac{2\pi}{\lambda} \mathbf{f}^T(\theta_{k,b}^\text{L,e} ,\phi_{k,b}^\text{L,e} ) \mathbf{r}_{b,n_x,n_y}^\text{S,L}  } h_{k,b,n_x,n_y}^\text{S,L} \nonumber\\
			&=\frac{1}{N} \sum_\iota \Lambda_{k,\iota,b} \eta_{k,\iota,b} \sum_{n_x} e^{ j \frac{2\pi}{\lambda} \left[ n_x d_\text{S} \left( -f_1(\theta_{k,b}^\text{L,e},\phi_{k,b}^\text{L,e}) + f_1(\theta_{k,\iota,b}^\text{L},\phi_{k,\iota,b}^\text{L}) \right)     + \mathcal{X}_1   \right]  } 
			+\frac{1}{N} \sum_\iota \Lambda_{k,\iota,b} \eta_{k,\iota,b} \sum_{n_x} e^{ j \frac{2\pi}{\lambda} \left[ n_x d_\text{S} \left( -f_1(\theta_{k,b}^\text{L,e},\phi_{k,b}^\text{L,e}) + f_1(\theta_{k,\iota,b}^\text{L},\phi_{k,\iota,b}^\text{L}) \right)  + \mathcal{X}_2  \right]  } \nonumber\\
			&+\frac{1}{N} \sum_\iota \Lambda_{k,\iota,b} \eta_{k,\iota,b} \sum_{n_y} e^{ j \frac{2\pi}{\lambda} \left[ \mathcal{X}_3  +  n_y d_\text{S} \left( -f_2(\theta_{k,b}^\text{L,e},\phi_{k,b}^\text{L,e}) + f_2(\theta_{k,\iota,b}^\text{L},\phi_{k,\iota,b}^\text{L}) \right)  \right]  } 
			+\frac{1}{N} \sum_\iota \Lambda_{k,\iota,b} \eta_{k,\iota,b} \sum_{n_y} e^{ j \frac{2\pi}{\lambda} \left[ \mathcal{X}_4  + n_y d_\text{S} \left( -f_2(\theta_{k,b}^\text{L,e},\phi_{k,b}^\text{L,e}) + f_2(\theta_{k,\iota,b}^\text{L},\phi_{k,\iota,b}^\text{L}) \right)  \right]  } \nonumber\\
			&-	\frac{1}{N} \sum_\iota \Lambda_{k,\iota,b} \eta_{k,\iota,b} e^{j \frac{2\pi}{\lambda} \mathcal{X}_1} - \frac{1}{N} \sum_\iota \Lambda_{k,\iota,b} \eta_{k,\iota,b} e^{j \frac{2\pi}{\lambda} \mathcal{X}_2} - \frac{1}{N} \sum_\iota \Lambda_{k,\iota,b} \eta_{k,\iota,b} e^{j \frac{2\pi}{\lambda} \mathcal{X}_3} - \frac{1}{N} \sum_\iota \Lambda_{k,\iota,b} \eta_{k,\iota,b} e^{j \frac{2\pi}{\lambda} \mathcal{X}_4}    \nonumber\\
			&=\frac{1}{N} \sum_\iota \Lambda_{k,\iota,b} \eta_{k,\iota,b} e^{ j \frac{2\pi}{\lambda} \left( \mathcal{X}_1 + \mathcal{X}_2 \right)  }  e^{ j \frac{\pi}{\lambda} (N_x-1)d_\text{S}\left( -f_1(\theta_{k,b}^\text{L,e},\phi_{k,b}^\text{L,e}) + f_1(\theta_{k,\iota,b}^\text{L},\phi_{k,\iota,b}^\text{L}) \right)  }  \frac{ \sin\frac{\pi}{\lambda} N_xd_\text{S}\left( -f_1(\theta_{k,b}^\text{L,e},\phi_{k,b}^\text{L,e}) + f_1(\theta_{k,\iota,b}^\text{L},\phi_{k,\iota,b}^\text{L}) \right)  }{ \sin\frac{\pi}{\lambda}d_\text{S} \left( -f_1(\theta_{k,b}^\text{L,e},\phi_{k,b}^\text{L,e}) + f_1(\theta_{k,\iota,b}^\text{L},\phi_{k,\iota,b}^\text{L}) \right) } \nonumber\\
%			&+\frac{1}{N} \sum_\iota \Lambda_{k,\iota,b} \eta_{k,\iota,b} e^{ j \frac{2\pi}{\lambda} \mathcal{X}_2  }  e^{ j \frac{\pi}{\lambda} (N_x-1)\left( -f_1(\theta_k^\text{L,e},\phi_k^\text{L,e}) + f_1(\theta_{k,\iota}^\text{L},\phi_{k,\iota}^\text{L}) \right)  }  \frac{ \sin\frac{\pi}{\lambda} N_x\left( -f_1(\theta_k^\text{L,e},\phi_k^\text{L,e}) + f_1(\theta_{k,\iota}^\text{L},\phi_{k,\iota}^\text{L}) \right)  }{ \sin\frac{\pi}{\lambda} \left( -f_1(\theta_k^\text{L,e},\phi_k^\text{L,e}) + f_1(\theta_{k,\iota}^\text{L},\phi_{k,\iota}^\text{L}) \right) } \nonumber\\
			&+\frac{1}{N} \sum_\iota \Lambda_{k,\iota,b} \eta_{k,\iota,b} e^{ j \frac{2\pi}{\lambda} \left( \mathcal{X}_3 + \mathcal{X}_4  \right) }  e^{ j \frac{\pi}{\lambda} (N_y-1)d_\text{S}\left( -f_2(\theta_{k,b}^\text{L,e},\phi_{k,b}^\text{L,e}) + f_2(\theta_{k,\iota,b}^\text{L},\phi_{k,\iota,b}^\text{L}) \right)  }  \frac{ \sin\frac{\pi}{\lambda} N_yd_\text{S}\left( -f_2(\theta_{k,b}^\text{L,e},\phi_{k,b}^\text{L,e}) + f_2(\theta_{k,\iota,b}^\text{L},\phi_{k,\iota,b}^\text{L}) \right)  }{ \sin\frac{\pi}{\lambda}d_\text{S} \left( -f_2(\theta_{k,b}^\text{L,e},\phi_{k,b}^\text{L,e}) + f_2(\theta_{k,\iota,b}^\text{L},\phi_{k,\iota,b}^\text{L}) \right) } \nonumber\\
%			&+\frac{1}{N} \sum_\iota \Lambda_{k,\iota,b} \eta_{k,\iota,b} e^{ j \frac{2\pi}{\lambda} \mathcal{X}_4  }  e^{ j \frac{\pi}{\lambda} (N_y-1)\left( -f_2(\theta_k^\text{L,e},\phi_k^\text{L,e}) + f_2(\theta_{k,\iota}^\text{L},\phi_{k,\iota}^\text{L}) \right)  }  \frac{ \sin\frac{\pi}{\lambda} N_y\left( -f_2(\theta_k^\text{L,e},\phi_k^\text{L,e}) + f_2(\theta_{k,\iota}^\text{L},\phi_{k,\iota}^\text{L}) \right)  }{ \sin\frac{\pi}{\lambda} \left( -f_2(\theta_k^\text{L,e},\phi_k^\text{L,e}) + f_2(\theta_{k,\iota}^\text{L},\phi_{k,\iota}^\text{L}) \right) } \nonumber\\
			&-	\frac{1}{N} \sum_\iota \Lambda_{k,\iota,b} \eta_{k,\iota,b} e^{j \frac{2\pi}{\lambda} \mathcal{X}_1} - \frac{1}{N} \sum_\iota \Lambda_{k,\iota,b} \eta_{k,\iota,b} e^{j \frac{2\pi}{\lambda} \mathcal{X}_2} - \frac{1}{N} \sum_\iota \Lambda_{k,\iota,b} \eta_{k,\iota,b} e^{j \frac{2\pi}{\lambda} \mathcal{X}_3} - \frac{1}{N} \sum_\iota \Lambda_{k,\iota,b} \eta_{k,\iota,b} e^{j \frac{2\pi}{\lambda} \mathcal{X}_4}    \nonumber\\
			&\overset{N_x,N_y \rightarrow \infty}{=} 
			\begin{cases}
				\Lambda_{k,\iota,b} \eta_{k,\iota,b},\text{ if } f_1(\theta_{k,b}^\text{L,e},\phi_{k,b}^\text{L,e}) = f_1(\theta_{k,\iota,b}^\text{L},\phi_{k,\iota,b}^\text{L}),\ f_2(\theta_{k,b}^\text{L,e},\phi_{k,b}^\text{L,e}) = f_2(\theta_{k,\iota,b}^\text{L},\phi_{k,\iota,b}^\text{L}), \\
				\ \ \ \ \ 0\ \ \ \ \ ,\text{ else}.
			\end{cases}	\label{eq71}
		\end{align}
		\rule{\linewidth}{1pt}
	\end{figure*}

	Then, the limitation of $\lim_{N\rightarrow\infty}\frac{1}{N} \sum_{n_x,n_y} e^{ j \frac{2\pi}{\lambda} \mathbf{f}^T(\theta_{k,b}^\text{L,e} ,\phi_{k,b}^\text{L,e} ) \mathbf{r}_{b,n_x,n_y}^\text{S,L}  } h_{k,b,n_x,n_y}^\text{S,L*} \left( s_{k,b,n_x,n_y}^\text{ref} \right)^2$ is expressed as Eq. \eqref{eq75}, where $\mathcal{Y}_1$, $\mathcal{Y}_2$, $\mathcal{Y}_3$ and $\mathcal{Y}_4$ are given as
		\begin{align}
		\mathcal{Y}_1  =&   \frac{-N_x+1}{2}d_\text{S} \left( f_1(\theta_{k,b}^\text{L,e},\phi_{k,b}^\text{L,e}) + f_1(\theta_{k,\iota,b}^\text{L},\phi_{k,\iota,b}^\text{L}) \right) \nonumber\\
		&+ \frac{-N_y+1}{2}d_\text{S} \left( f_2(\theta_{k,b}^\text{L,e},\phi_{k,b}^\text{L,e}) + f_2(\theta_{k,\iota,b}^\text{L},\phi_{k,\iota,b}^\text{L}) \right)  ,\\
	\mathcal{Y}_2  =&   \frac{-N_x+1}{2}d_\text{S} \left( f_1(\theta_{k,b}^\text{L,e},\phi_{k,b}^\text{L,e}) + f_1(\theta_{k,\iota,b}^\text{L},\phi_{k,\iota,b}^\text{L}) \right) \nonumber\\
	&+ \frac{N_y-1}{2}d_\text{S} \left( f_2(\theta_{k,b}^\text{L,e},\phi_{k,b}^\text{L,e}) + f_2(\theta_{k,\iota,b}^\text{L},\phi_{k,\iota,b}^\text{L}) \right) ,\\
	\mathcal{Y}_3  =&   \frac{N_x-1}{2}d_\text{S} \left( f_1(\theta_{k,b}^\text{L,e},\phi_{k,b}^\text{L,e}) + f_1(\theta_{k,\iota,b}^\text{L},\phi_{k,\iota,b}^\text{L}) \right) \nonumber\\
	&+ \frac{-N_y+1}{2}d_\text{S} \left( f_2(\theta_{k,b}^\text{L,e},\phi_{k,b}^\text{L,e}) + f_2(\theta_{k,\iota,b}^\text{L},\phi_{k,\iota,b}^\text{L}) \right) ,\\
	\mathcal{Y}_4  =&   \frac{N_x-1}{2}d_\text{S} \left( f_1(\theta_{k,b}^\text{L,e},\phi_{k,b}^\text{L,e}) + f_1(\theta_{k,\iota,b}^\text{L},\phi_{k,\iota,b}^\text{L}) \right) \nonumber\\
	&+ \frac{N_y-1}{2}d_\text{S} \left( f_2(\theta_{k,b}^\text{L,e},\phi_{k,b}^\text{L,e}) + f_2(\theta_{k,\iota,b}^\text{L},\phi_{k,\iota,b}^\text{L}) \right).
\end{align}		
	
\begin{figure*}
\begin{align}
	&\frac{1}{N} \sum_{n_x,n_y} e^{ - j \frac{2\pi}{\lambda} \mathbf{f}^T(\theta_{k,b}^\text{L,e} ,\phi_{k,b}^\text{L,e} ) \mathbf{r}_{b,n_x,n_y}^\text{S,L}  } h_{k,b,n_x,n_y}^\text{S,L*} \left( s_{k,b,n_x,n_y}^\text{ref} \right)^2 \nonumber\\
	&=\frac{A}{N} \sum_{n_x,n_y} e^{ j \frac{2\pi}{\lambda} \mathbf{f}^T(\theta_{k,b}^\text{L,e} ,\phi_{k,b}^\text{L,e} ) \mathbf{r}_{b,n_x,n_y}^\text{S,L}  } h_{k,b,n_x,n_y}^\text{S,L*} e^{ j 2 \pi \chi_{k,b,n_x,n_y} } \nonumber\\
	&=\frac{A}{N} \sum_\iota \Lambda_{k,\iota,b} \eta_{k,\iota,b} \sum_{n_x} e^{ -j \frac{2\pi}{\lambda} \left[ n_x d_\text{S} \left( f_1(\theta_{k,b}^\text{L,e},\phi_{k,b}^\text{L,e}) + f_1(\theta_{k,\iota,b}^\text{L},\phi_{k,\iota,b}^\text{L}) \right)  + \mathcal{Y}_1  \right]  } e^{ j 2 \pi \chi_{k,b,n_x,0} } \nonumber\\
	&\ \ \  + \frac{A}{N} \sum_\iota \Lambda_{k,\iota,b} \eta_{k,\iota,b} \sum_{n_x} e^{ -j \frac{2\pi}{\lambda} \left[ n_x d_\text{S} \left( f_1(\theta_{k,b}^\text{L,e},\phi_{k,b}^\text{L,e}) + f_1(\theta_{k,\iota,b}^\text{L},\phi_{k,\iota,b}^\text{L}) \right)  + \mathcal{Y}_2  \right]  } e^{ j 2 \pi \chi_{k,b,n_x,N_y-1} }  \nonumber\\
	&\ \ \ + \frac{A}{N} \sum_\iota \Lambda_{k,\iota,b} \eta_{k,\iota,b} \sum_{n_y} e^{ -j \frac{2\pi}{\lambda} \left[ \mathcal{Y}_3 + n_y d_\text{S} \left( f_2(\theta_{k,b}^\text{L,e},\phi_{k,b}^\text{L,e}) + f_2(\theta_{k,\iota,b}^\text{L},\phi_{k,\iota,b}^\text{L}) \right)    \right]  } e^{ j 2 \pi \chi_{k,b,0,n_y} }  \nonumber\\
	&\ \ \ + \frac{A}{N} \sum_\iota \Lambda_{k,\iota,b} \eta_{k,\iota,b} \sum_{n_y} e^{ -j \frac{2\pi}{\lambda} \left[ \mathcal{Y}_4 + n_y d_\text{S} \left( f_2(\theta_{k,b}^\text{L,e},\phi_{k,b}^\text{L,e}) + f_2(\theta_{k,\iota,b}^\text{L},\phi_{k,\iota,b}^\text{L}) \right)    \right]  } e^{ j 2 \pi \chi_{k,b,0,N_y-1} }  \nonumber\\
	&\ \ \ -	\frac{A}{N} \sum_\iota \Lambda_{k,\iota,b} \eta_{k,\iota,b} e^{-j \frac{2\pi}{\lambda} \mathcal{Y}_1} e^{ j 2 \pi \chi_{k,b,0,0} }  - \frac{A}{N} \sum_\iota \Lambda_{k,\iota,b} \eta_{k,\iota,b} e^{-j \frac{2\pi}{\lambda} \mathcal{Y}_2} e^{ j 2 \pi \chi_{k,b,0,N_y-1} }       \nonumber\\
	&\ \ \ - \frac{A}{N} \sum_\iota \Lambda_{k,\iota,b} \eta_{k,\iota,b} e^{-j \frac{2\pi}{\lambda} \mathcal{Y}_3} e^{ j 2 \pi \chi_{k,b,N_x-1,0} }  - \frac{A}{N} \sum_\iota \Lambda_{k,\iota,b} \eta_{k,\iota,b} e^{-j \frac{2\pi}{\lambda} \mathcal{Y}_4} e^{ j 2 \pi \chi_{k,b,N_x-1,N_y-1} } \nonumber\\
	&\overset{N\rightarrow \infty}{=} \frac{A}{N} \sum_\iota \Lambda_{k,\iota,b}  \eta_{k,\iota,b} \mathbb{E} \left[  e^{ j 2 \pi \chi_{k,b,n_x-1,n_y-1} } \right] \nonumber\\
	& = 0.		\label{eq75}
\end{align}
\rule{\linewidth}{1pt}
\end{figure*}	
	
		Similar to Eq. \eqref{eq71} and Eq. \eqref{eq75}, we have 
	\begin{align}
		&\lim_{N\rightarrow\infty}\frac{1}{N} \sum_{n_x,n_y} e^{ - j \frac{2\pi}{\lambda} \mathbf{f}^T(\theta_{k,b}^\text{L,e} ,\phi_{k,b}^\text{L,e} ) \mathbf{r}_{b,n_x,n_y}^\text{S,L}  } z_{k,b,n_x,n_y}^\text{S} = 0, \label{eq80}\\
		&\lim_{N\rightarrow\infty}\frac{1}{N} \sum_{n_x,n_y} e^{ - j \frac{2\pi}{\lambda} \mathbf{f}^T(\theta_{k,b}^\text{L,e} ,\phi_{k,b}^\text{L,e} ) \mathbf{r}_{b,n_x,n_y}^\text{S,L}  } z_{k,b,n_x,n_y}^\text{S*} \left( s_{k,b,n_x,n_y}^\text{ref} \right)^2=0. \label{eq81}
	\end{align}
	
	In conclusion, according to Eq. \eqref{eq71}, Eq. \eqref{eq75}, Eq. \eqref{eq80} and Eq. \eqref{eq81}, we have 
		\begin{align}
	\tilde{\mathcal{H}}_{k,b}  =
	\begin{cases}
		\sqrt{P^\text{S}} A \Lambda_{k,\iota,b}\eta_{k,\iota,b},\text{ if }f_1(\theta_{k,b}^\text{L,e},\phi_{k,b}^\text{L,e}) = f_1(\theta_{k,\iota,b}^\text{L},\phi_{k,\iota,b}^\text{L})\ \text{and}\\ \ \ \ \ \ \ \ \ \ \ \ \ \ \ \ \ f_2(\theta_{k,b}^\text{L,e},\phi_{k,b}^\text{L,e}) = f_2(\theta_{k,\iota,b}^\text{L},\phi_{k,\iota,b}^\text{L}),\\
		0,\text{ else}.
	\end{cases}  
\end{align}
The proof of Theorem 1 is completed.

\section{Proof of Lemma 2}
Denote $(\theta_0^\text{L},\phi_0^\text{L})$ and $(\theta^\text{L,e},\phi^\text{L,e})$ as the true angles and the estimated angles in the local coordinate system, which satisfies $-\pi<\theta_0^\text{L},\theta^\text{L,e}\leq\pi$ and $-\frac{\pi}{2}<\phi_0^\text{L},\phi^\text{L,e}\leq\frac{\pi}{2}$. The ambiguity function is given as
\begin{align}
	&\left| \frac{1}{N} \sum_{n_x,n_y} e^{ -j \frac{2\pi}{\lambda} \mathbf{f}^T(\theta^\text{L,e} ,\phi^\text{L,e} ) \mathbf{r}_{b,n_x,n_y}^\text{S,L}  } e^{ j \frac{2\pi}{\lambda} \mathbf{f}^T(\theta_{0}^\text{L} ,\phi_{0}^\text{L} ) \mathbf{r}_{b,n_x,n_y}^\text{S,L}  } \right| \nonumber\\
	&= \left| \frac{1}{N}  \sum_{n_x} e^{ j \frac{2\pi}{\lambda} \left[ n_x d_\text{S} \left( -f_1(\theta^\text{L,e},\phi^\text{L,e}) + f_1(\theta_0^\text{L},\phi_0^\text{L}) \right)     + \mathcal{Z}_1 + \mathcal{Z}_2  \right]  } \right. \nonumber\\
	& \ \ \ + \frac{1}{N}  \sum_{n_y} e^{ j \frac{2\pi}{\lambda} \left[ n_y d_\text{S} \left( -f_2(\theta^\text{L,e},\phi^\text{L,e}) + f_2(\theta_0^\text{L},\phi_0^\text{L}) \right)     + \mathcal{Z}_3 + \mathcal{Z}_4  \right]  } \nonumber\\
	&\ \ \ \left. - \frac{1}{N} e^{ j \frac{2\pi}{\lambda} \mathcal{Z}_1 } - \frac{1}{N} e^{ j \frac{2\pi}{\lambda} \mathcal{Z}_2 } - - \frac{1}{N} e^{ j \frac{2\pi}{\lambda} \mathcal{Z}_3 } - - \frac{1}{N} e^{ j \frac{2\pi}{\lambda} \mathcal{Z}_4 } \right|. \label{eq83}
\end{align}
where we have
	\begin{align}
	\mathcal{Z}_1  =&   \frac{-N_x+1}{2}d_\text{S} \left( -f_1(\theta^\text{L,e},\phi^\text{L,e}) + f_1(\theta_0^\text{L},\phi_0^\text{L}) \right) \nonumber\\
	&+ \frac{-N_y+1}{2}d_\text{S} \left( -f_2(\theta^\text{L,e},\phi^\text{L,e}) + f_2(\theta_0^\text{L},\phi_0^\text{L}) \right) ,\\
	\mathcal{Z}_2  =&   \frac{-N_x+1}{2}d_\text{S} \left( -f_1(\theta^\text{L,e},\phi^\text{L,e}) + f_1(\theta_{0}^\text{L},\phi_{0}^\text{L}) \right) \nonumber\\
	&+ \frac{N_y-1}{2}d_\text{S} \left( -f_2(\theta^\text{L,e},\phi^\text{L,e}) + f_2(\theta_{0}^\text{L},\phi_{0}^\text{L}) \right),\\
	\mathcal{Z}_3  =&   \frac{N_x-1}{2}d_\text{S} \left( -f_1(\theta^\text{L,e},\phi^\text{L,e}) + f_1(\theta_{0}^\text{L},\phi_{0}^\text{L}) \right) \nonumber\\
	& +  \frac{-N_y+1}{2}d_\text{S} \left( -f_2(\theta^\text{L,e},\phi^\text{L,e}) + f_2(\theta_{0}^\text{L},\phi_{0}^\text{L}) \right) ,\\
	\mathcal{Z}_4  =  & \frac{N_x-1}{2}d_\text{S} \left( -f_1(\theta^\text{L,e},\phi^\text{L,e}) + f_1(\theta_{0}^\text{L},\phi_{0}^\text{L}) \right) \nonumber\\
	& + \frac{N_y-1}{2}d_\text{S} \left( -f_2(\theta^\text{L,e},\phi^\text{L,e}) + f_2(\theta_{0}^\text{L},\phi_{0}^\text{L}) \right) .
\end{align}
The ambiguity function \eqref{eq83} achieves the maximal value if and only if $ -f_1(\theta^\text{L,e},\phi^\text{L,e}) + f_1(\theta_0^\text{L},\phi_0^\text{L}) =  p^\prime \frac{\lambda}{d_\text{S}}$ and $-f_2(\theta^\text{L,e},\phi^\text{L,e}) + f_2(\theta_0^\text{L},\phi_0^\text{L})=q^\prime \frac{\lambda}{d_\text{S}}$, where $p^\prime, q^\prime \in \mathbb{Z}$.
Observing Eq. \eqref{eq83}, if we want to correctly estimate the direction vector $\mathbf{f}(\theta_0^\text{L},\phi_0^\text{L})$ by adopting the FFT-based detection method and avoid the angle aliasing, we have $-1 < p^\prime < 1 $ and $-1 < q^\prime < 1 $, while the following conditions should be satisfied
\begin{align}
	&\left| \frac{ d_\text{S} \left( -f_1(\theta^\text{L,e},\phi^\text{L,e}) + f_1(\theta_0^\text{L},\phi_0^\text{L}) \right) }{\lambda} \right| < 1, \label{eq84}\\
	&\left| \frac{ d_\text{S} \left( -f_2(\theta^\text{L,e},\phi^\text{L,e}) + f_2(\theta_0^\text{L},\phi_0^\text{L}) \right) }{\lambda} \right| < 1. \label{eq85}
\end{align}
Since we have $\left| -f_1(\theta^\text{L,e},\phi^\text{L,e}) + f_1(\theta_0^\text{L},\phi_0^\text{L}) \right| < 2$ and $\left| -f_2(\theta^\text{L,e},\phi^\text{L,e}) + f_2(\theta_0^\text{L},\phi_0^\text{L}) \right| < 2$, the constraints \eqref{eq84} and \eqref{eq85} can be reformulated as
\begin{align}
	&  d_\text{S}   <  \frac{\lambda}{\left| -f_1(\theta^\text{L,e},\phi^\text{L,e}) + f_1(\theta_0^\text{L},\phi_0^\text{L}) \right|},\label{eq86}\\
	&d_\text{S}   <  \frac{\lambda}{\left| -f_2(\theta^\text{L,e},\phi^\text{L,e}) + f_2(\theta_0^\text{L},\phi_0^\text{L}) \right|}. \label{eq87}
\end{align}
Moreover, we have
\begin{align}
	&\frac{\lambda}{\left| -f_1(\theta^\text{L,e},\phi^\text{L,e}) + f_1(\theta_0^\text{L},\phi_0^\text{L}) \right|} >\frac{\lambda}{2}, \label{eq88} \\
	&\frac{\lambda}{\left| -f_2(\theta^\text{L,e},\phi^\text{L,e}) + f_2(\theta_0^\text{L},\phi_0^\text{L}) \right|} >\frac{\lambda}{2}. \label{eq89}
\end{align}
Finally, the constraints \eqref{eq84}-\eqref{eq89}, we have 
\begin{align}
	d_\text{S} \leq \frac{\lambda}{2}.
\end{align}
Therefore, arrange the sensing element with the half-wavelength spacing is sufficient and laying the sensing elements with sub-wavelength is not required. The proof of  \textit{Theorem 2} is completed.

\end{appendices}

\bibliographystyle{IEEEtran}
\bibliography{reference}

\begin{thebibliography}{10}
\providecommand{\url}[1]{#1}
\csname url@samestyle\endcsname
\providecommand{\newblock}{\relax}
\providecommand{\bibinfo}[2]{#2}
\providecommand{\BIBentrySTDinterwordspacing}{\spaceskip=0pt\relax}
\providecommand{\BIBentryALTinterwordstretchfactor}{4}
\providecommand{\BIBentryALTinterwordspacing}{\spaceskip=\fontdimen2\font plus
\BIBentryALTinterwordstretchfactor\fontdimen3\font minus
  \fontdimen4\font\relax}
\providecommand{\BIBforeignlanguage}[2]{{%
\expandafter\ifx\csname l@#1\endcsname\relax
\typeout{** WARNING: IEEEtran.bst: No hyphenation pattern has been}%
\typeout{** loaded for the language `#1'. Using the pattern for}%
\typeout{** the default language instead.}%
\else
\language=\csname l@#1\endcsname
\fi
#2}}
\providecommand{\BIBdecl}{\relax}
\BIBdecl

\bibitem{10892042}
K.~Ahmadi and W.~A. Serdijn, ``Advancements in laser and led-based optical
  wireless power transfer for iot applications: A comprehensive review,''
  \emph{IEEE Internet of Things Journal}, pp. 1--1, 2025.

\bibitem{10430083}
Y.~Zheng, J.~Hu, Y.~Zhao, and K.~Yang, ``Average age of sensing in wireless
  powered sensor networks,'' \emph{IEEE Transactions on Wireless
  Communications}, vol.~23, no.~8, pp. 9265--9281, 2024.

\bibitem{10663809}
X.~Li, Z.~Han, G.~Zhu, Y.~Shi, J.~Xu, Y.~Gong, Q.~Zhang, K.~Huang, and K.~B.
  Letaief, ``Integrating sensing, communication, and power transfer: From
  theory to practice,'' \emph{IEEE Communications Magazine}, vol.~62, no.~9,
  pp. 122--127, 2024.

\bibitem{10599126}
R.~Ma, J.~Tang, X.~Y. Zhang, W.~Feng, D.~K.~C. So, C.-B. Chae, K.-K. Wong, and
  J.~A. Chambers, ``Simultaneous wireless information and power transfer in
  iot-based scenarios: Architectures, challenges, and prototype validation,''
  \emph{IEEE Wireless Communications}, vol.~31, no.~5, pp. 272--278, 2024.

\bibitem{9454368}
G.~Kwon, H.~Park, and M.~Z. Win, ``Joint beamforming and power splitting for
  wideband millimeter wave swipt systems,'' \emph{IEEE Journal of Selected
  Topics in Signal Processing}, vol.~15, no.~5, pp. 1211--1227, 2021.

\bibitem{7350112}
J.~S. Herd and M.~D. Conway, ``The evolution to modern phased array
  architectures,'' \emph{Proceedings of the IEEE}, vol. 104, no.~3, pp.
  519--529, 2016.

\bibitem{9696209}
R.~Deng, B.~Di, H.~Zhang, Y.~Tan, and L.~Song, ``Reconfigurable holographic
  surface-enabled multi-user wireless communications: Amplitude-controlled
  holographic beamforming,'' \emph{IEEE Transactions on Wireless
  Communications}, vol.~21, no.~8, pp. 6003--6017, 2022.

\bibitem{Shao20246DMA}
\BIBentryALTinterwordspacing
X.~Shao, R.~Zhang, Q.~Jiang, and R.~Schober, ``6d movable antenna enhanced
  wireless network via discrete position and rotation optimization,''
  \emph{ArXiv}, vol. abs/2403.17122, 2024. [Online]. Available:
  \url{https://api.semanticscholar.org/CorpusID:268691662}
\BIBentrySTDinterwordspacing

\bibitem{4595260}
L.~R. Varshney, ``Transporting information and energy simultaneously,'' in
  \emph{2008 IEEE International Symposium on Information Theory}, 2008, pp.
  1612--1616.

\bibitem{8115220}
B.~Clerckx, ``Wireless information and power transfer: Nonlinearity, waveform
  design, and rate-energy tradeoff,'' \emph{IEEE Transactions on Signal
  Processing}, vol.~66, no.~4, pp. 847--862, 2018.

\bibitem{11030755}
S.~Na, F.~Hu, Z.~Ling, H.~Dong, and X.~Yao, ``Joint frequency-time allocation
  and phase-shift optimization in intelligent reflecting surface assisted
  multigroup wpcn,'' \emph{IEEE Internet of Things Journal}, vol.~12, no.~17,
  pp. 35\,249--35\,260, 2025.

\bibitem{9681843}
R.~Deng, B.~Di, H.~Zhang, and L.~Song, ``Hdma: Holographic-pattern division
  multiple access,'' \emph{IEEE Journal on Selected Areas in Communications},
  vol.~40, no.~4, pp. 1317--1332, 2022.

\bibitem{9393594}
B.~Di, ``Reconfigurable holographic metasurface aided wideband ofdm
  communications against beam squint,'' \emph{IEEE Transactions on Vehicular
  Technology}, vol.~70, no.~5, pp. 5099--5103, 2021.

\bibitem{azarbahram2024waveformoptimizationbeamfocusing}
\BIBentryALTinterwordspacing
A.~Azarbahram, O.~L.~A. López, and M.~Latva-Aho, ``Waveform optimization and
  beam focusing for near-field wireless power transfer with dynamic metasurface
  antennas and non-linear energy harvesters,'' 2024. [Online]. Available:
  \url{https://arxiv.org/abs/2307.01081}
\BIBentrySTDinterwordspacing

\bibitem{10721321}
Q.~Huang, J.~Hu, Y.~Zhao, and K.~Yang, ``Holographic integrated data and energy
  transfer,'' \emph{IEEE Transactions on Wireless Communications}, vol.~23,
  no.~12, pp. 18\,987--19\,002, 2024.

\bibitem{9264694}
K.-K. Wong, A.~Shojaeifard, K.-F. Tong, and Y.~Zhang, ``Fluid antenna
  systems,'' \emph{IEEE Transactions on Wireless Communications}, vol.~20,
  no.~3, pp. 1950--1962, 2021.

\bibitem{10980171}
X.~Lin, Y.~Zhao, H.~Yang, J.~Hu, and K.-K. Wong, ``Fluid antenna multiple
  access assisted integrated data and energy transfer: Outage and multiplexing
  gain analysis,'' \emph{IEEE Transactions on Wireless Communications}, pp.
  1--1, 2025.

\bibitem{10752873}
X.~Shao, Q.~Jiang, and R.~Zhang, ``6d movable antenna based on user
  distribution: Modeling and optimization,'' \emph{IEEE Transactions on
  Wireless Communications}, pp. 1--1, 2024.

\bibitem{111111}
W.~Zhonglun, Z.~Yizhe, H.~Jie, and Y.~Kun, ``6dma-assisted integrated data and
  energy transfer: Joint spatial orientation and beamforming design,'' in
  \emph{2025 IEEE 102th Vehicular Technology Conference (VTC2025-Fall)}, 2025,
  pp. 1--1.

\bibitem{Shao2024DistributedCE}
\BIBentryALTinterwordspacing
X.~Shao, R.~Zhang, Q.~Jiang, J.~Park, T.~Q. S.Quek, and R.~Schober,
  ``Distributed channel estimation and optimization for 6d movable antenna:
  Unveiling directional sparsity,'' 2024. [Online]. Available:
  \url{https://api.semanticscholar.org/CorpusID:272880808}
\BIBentrySTDinterwordspacing

\bibitem{10158988}
H.~Li, S.~Shen, and B.~Clerckx, ``Beyond diagonal reconfigurable intelligent
  surfaces: A multi-sector mode enabling highly directional full-space wireless
  coverage,'' \emph{IEEE Journal on Selected Areas in Communications}, vol.~41,
  no.~8, pp. 2446--2460, 2023.

\bibitem{9724245}
H.~Zhang, H.~Zhang, B.~Di, M.~D. Renzo, Z.~Han, H.~V. Poor, and L.~Song,
  ``Holographic integrated sensing and communication,'' \emph{IEEE Journal on
  Selected Areas in Communications}, vol.~40, no.~7, pp. 2114--2130, 2022.

\bibitem{9795244}
J.~Hu, Y.~Zheng, and K.~Yang, ``Multi-domain resource scheduling for
  full-duplex aided wireless powered communication network,'' \emph{IEEE
  Transactions on Vehicular Technology}, vol.~71, no.~10, pp. 10\,849--10\,862,
  2022.

\bibitem{10163760}
R.~Deng, Y.~Zhang, H.~Zhang, B.~Di, H.~Zhang, H.~V. Poor, and L.~Song,
  ``Reconfigurable holographic surfaces for ultra-massive mimo in 6g: Practical
  design, optimization and implementation,'' \emph{IEEE Journal on Selected
  Areas in Communications}, vol.~41, no.~8, pp. 2367--2379, 2023.

\bibitem{9848831}
R.~Deng, B.~Di, H.~Zhang, H.~V. Poor, and L.~Song, ``Holographic mimo for leo
  satellite communications aided by reconfigurable holographic surfaces,''
  \emph{IEEE Journal on Selected Areas in Communications}, vol.~40, no.~10, pp.
  3071--3085, 2022.

\bibitem{9950543}
Y.~Zhao, Y.~Wu, J.~Hu, and K.~Yang, ``A general analysis and optimization
  framework of time index modulation for integrated data and energy transfer,''
  \emph{IEEE Transactions on Wireless Communications}, vol.~22, no.~6, pp.
  3657--3670, 2023.

\bibitem{10845870}
S.~Liu, X.~Yu, Z.~Gao, J.~Xu, D.~W.~K. Ng, and S.~Cui, ``Sensing-enhanced
  channel estimation for near-field xl-mimo systems,'' \emph{IEEE Journal on
  Selected Areas in Communications}, pp. 1--1, 2025.

\bibitem{Nocedal2006}
\BIBentryALTinterwordspacing
\emph{Interior-Point Methods for Nonlinear Programming}.\hskip 1em plus 0.5em
  minus 0.4em\relax New York, NY: Springer New York, 2006, pp. 563--597.
  [Online]. Available: \url{https://doi.org/10.1007/978-0-387-40065-5_19}
\BIBentrySTDinterwordspacing

\bibitem{10097166}
Q.~Yang, A.~Guerra, F.~Guidi, N.~Shlezinger, H.~Zhang, D.~Dardari, B.~Wang, and
  Y.~C. Eldar, ``Near-field localization with dynamic metasurface antennas,''
  in \emph{ICASSP 2023 - 2023 IEEE International Conference on Acoustics,
  Speech and Signal Processing (ICASSP)}, 2023, pp. 1--5.

\bibitem{8038776}
M.~Chen, Y.~Miao, Y.~Hao, and K.~Hwang, ``Narrow band internet of things,''
  \emph{IEEE Access}, vol.~5, pp. 20\,557--20\,577, 2017.

\bibitem{9400843}
C.~Hu, L.~Dai, S.~Han, and X.~Wang, ``Two-timescale channel estimation for
  reconfigurable intelligent surface aided wireless communications,''
  \emph{IEEE Transactions on Communications}, vol.~69, no.~11, pp. 7736--7747,
  2021.

\end{thebibliography}

\end{document}